\documentclass[10pt,a4paper]{article}

\usepackage{graphicx,comment,amsmath,listings,longtable}
\usepackage{url}
\newcommand{\ic}{\textrm{i}}
\newcommand{\Arg}{\textrm{Arg}}
\newcommand{\ok}{$\cdot$}
\DeclareMathOperator\arcsinh{arcsinh}
\DeclareMathOperator\arccosh{arccosh}
\DeclareMathOperator\arctanh{arctanh}

\title{On quality of implementation of Fortran 2008 complex
intrinsic functions on branch cuts}

\author{Anton Shterenlikht \\
Mech Eng Dept, The University of Bristol, Bristol BS8 1TR, UK}

\begin{document}

\maketitle

\lstset{basicstyle=\ttfamily}

\section*{Abstract}

Branch cuts in complex functions in combination
with signed zero and signed infinity have important
uses in fracture mechanics, jet flow and aerofoil
analysis.
We present benchmarks for validating
Fortran 2008 complex functions - LOG, SQRT,
ASIN, ACOS, ATAN, ASINH, ACOSH and ATANH -
on branch cuts with arguments
of all 3 IEEE floating point binary formats:
binary32, binary64 and binary128.
Results are reported with 8 Fortran 2008 compilers:
GCC, Flang, Cray, Oracle, PGI, Intel, NAG and IBM.
Multiple test failures were revealed, e.g.
wrong signs of results or
unexpected overflow, underflow, or NaN.
We conclude that the quality of implementation
of these Fortran 2008 intrinsics in many compilers
is not yet sufficient to remove
the need for special code for branch cuts.
The test results are complemented by conformal maps
of the branch cuts and detailed derivations
of the values of these functions
on branch cuts, to be used as a reference.
The benchmarks are freely available from
\url{cmplx.sf.net}.
This work will be of interest to engineers who
use complex functions, as well as to
compiler and maths library developers.

\section{Introduction}
\label{sec:intro}

In the following
$ w = u + \ic v $
and
$ z = x + \ic y $
are complex variables,
and
$ w = w ( z ) $
is a conformal mapping function from $ z $ to $ w $.
$ \Re z $ and $ \Im z $ are the real and the imaginary
parts of $ z $.
Fortran functions and named constants are written
in \texttt{monospaced} font.

Fortran intrinsic functions \texttt{SQRT} and \texttt{LOG} accepted
complex arguments at least since FORTRAN66 standard \cite{f66}.
IEEE floating point standard \cite{ieee754-1985}
defined signed zero and signed infinity:
$ + 0 , - 0 , + \infty , - \infty $.
Fortran 95 standard \cite{f95} added support
for IEEE floating point arithmetic.
Fortran 2008 standard \cite{f2008} added support
for complex arguments to intrinsic
functions \texttt{ACOS}, \texttt{ASIN}, \texttt{ATAN}
and 3 new inverse hyperbolic intrinsics:
\texttt{ACOSH}, \texttt{ASINH},
\texttt{ATANH}, all of which also accept complex arguments.

\begin{figure}[ht!]
\centering
\includegraphics[ width = 0.9 \textwidth ]{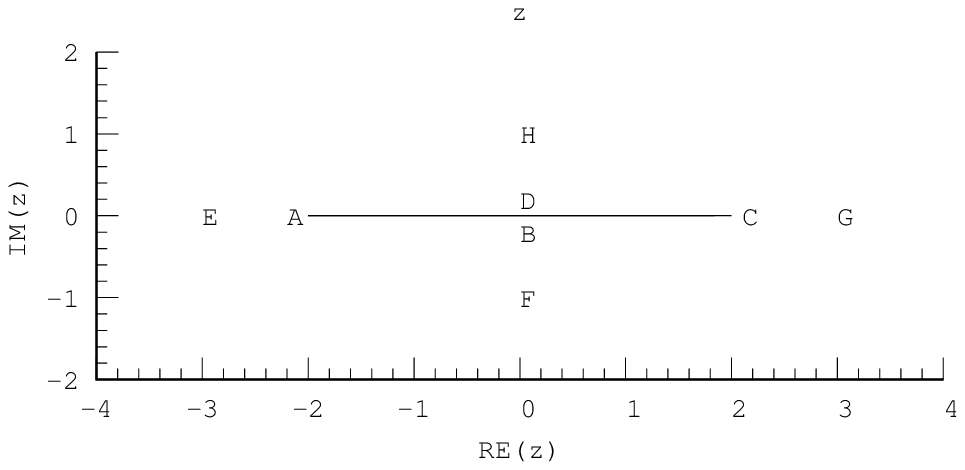}\\
\includegraphics[ width = 0.9 \textwidth ]{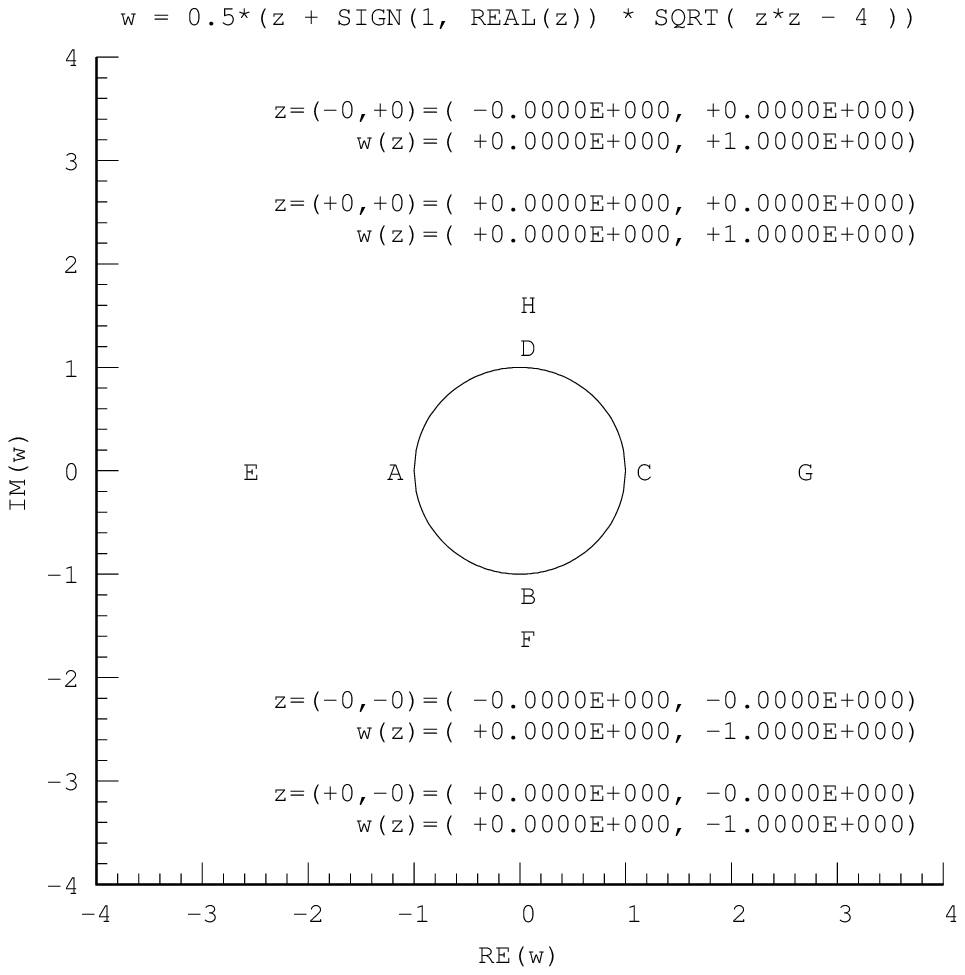}
\caption{Map of
$ w = \frac{1}{2}(z + \textrm{copySign}(1, \Re z ) \sqrt{z^2 - 4} ) $.
The branch cut ABCD in $ z $ is mapped onto a unit
circle ABCD in $ w $. }
\label{fig:muha}
\end{figure}

These 8 complex elementary functions, together
with IEEE style signed zero and infinity, have useful
applications e.g. in fracture mechanics, because
a branch cut can represent a mathematical crack.
Perhaps the oldest and simplest example is function

\begin{equation}
 z = w + \frac{1}{w}
\label{eq:loza}
\end{equation}
which maps a complex plane with a cut unit circle
onto a complex plane with a cut along $ x $ at $ - 2 \le x \le 2 $.
This function has been in use probably since early 20th century,
see e.g. \cite{kober1952, muskhelishvili1953}.
It is still widely used in fracture mechanics today
\cite{lopez2009}.
In practice the inverse of Eqn. \eqref{eq:loza} is more
useful:

\begin{equation}
 w = \frac{1}{2}(z + \textrm{copySign}(1,\Re z ) \sqrt{z^2 - 4} )
\label{eq:bober}
\end{equation}
where \emph{copySign} is an IEEE function which returns a value
with the magnitude of the first argument and the sign of
the second argument \cite{60559-2011}.
The map of Eqn. \eqref{eq:bober} is shown in Fig. \ref{fig:muha}

Note that Eqn. \eqref{eq:bober} produces the desired mapping
only if $ + 0 $ and $ - 0 $ can be distinguished,
so that points in $ z $ on the top and the bottom boundary
of the cut, i.e. with $ y = + 0 $ and $ y = - 0 $
are mapped respectively onto the top and the bottom boundary
of the unit circle in $ w $.
For example point
$ z = 0 - \ic 0 $
is mapped to point
$ w = 0 - \ic 1 $,
point B in Fig. \ref{fig:muha},
and point
$ z = 0 + \ic 0 $
is mapped to point
$ w = 0 + \ic 1 $,
point D in Fig. \ref{fig:muha}.

\begin{figure}[ht!]
\centering
\includegraphics[width= 0.9 \textwidth]{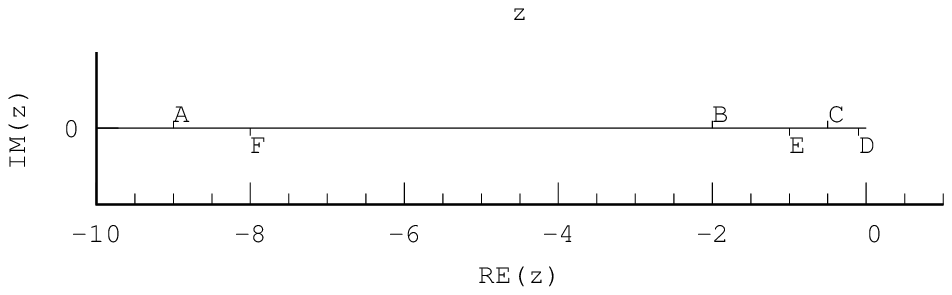} \\
\includegraphics[width= 0.9 \textwidth]{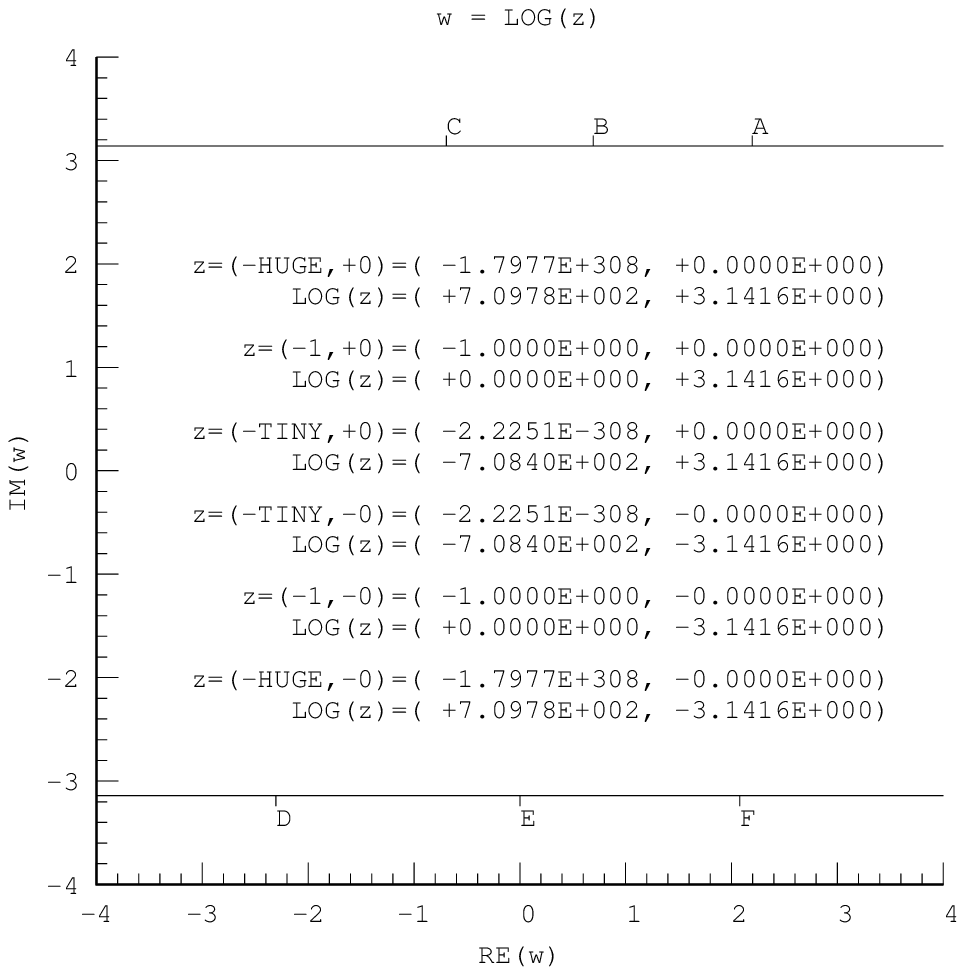}
\caption{ Map of the branch cut of $ w = \log z $.
Points A, B, C are on the top boundary of the cut, $ y = + 0 $.
Points D, E, F are on the bottom boundary of the cut, $ y = - 0 $. }
\label{fig:log1}
\end{figure}

$ \log z $ has a single branch cut along the negative
real axis, therefore it can be used for analysis
of an edge crack in an infinite plate.
A map of $ w = \log z $ is shown in Fig. \ref{fig:log1}.

The 3 inverse trigonometric
($ \arcsin $ , $ \arccos $ and $ \arctan $) and
the 2 inverse hyperbolic functions
($ \arcsinh $ and $ \arctanh $) have 2 cuts
on either the real or the imaginary axis,
and can therefore be used for the analysis
of bodies with 2 cracks along the same line,
e.g. an infinite or a finite width plate with
2 opposing cracks with a finite ligament length
in between.
This case is of significant practical importance
in fracture mechanics,
see e.g. \cite[Sec. 4, `Parallel Cracks']{tada2000}.

The third inverse hyperbolic function, $ \arccosh $,
has a single branch cut and can be used for
an edge crack geometry.

\begin{figure}[ht!]
\centering
\includegraphics[ width = 0.9 \textwidth ]{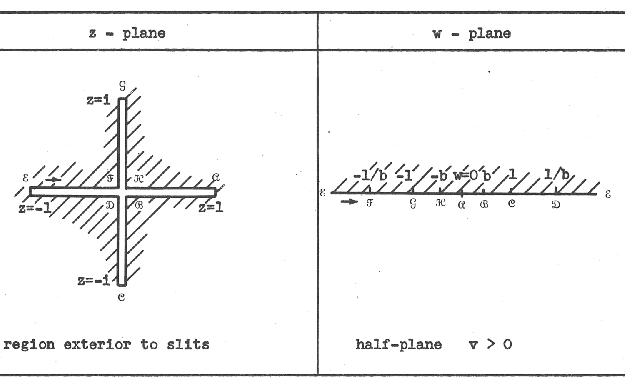}
\caption{Map of function $ w = \tan ( \arccos z^2 / 4 ) $,
reproduced from \cite{kober1952}. }
\label{fig:cross}
\end{figure}

Another interesting case is function
$ w = \tan ( \arccos z^2 / 4 ) $
\cite[p. 79]{kober1952}
which maps a plane with 2 intersecting cuts onto an upper half plane,
$ v \ge 0 $.
The two cuts form a cross centred at the origin,
see Fig. \ref{fig:cross}.

Jet flows and aerofoils
are among other popular practical examples
where signed zero is required to obtain correct
conformal maps of multivalued complex functions
on branch cuts
\cite{kahan87, kahan97}.

The usage of $ - 0 $ was further popularised,
although with no new examples,
in \cite{goldberg91, overton2001}. 

Although algorithms can be, and have been, developed
which use data a short distance away from the cuts,
this is not very satisfactory, as it is not an obvious
question what this small distance should be.
In addition, branch cuts often contain the most
important data, e.g. the extremum values of crack
tip displacement fields are found on crack flanks,
which is useful in experimental fracture mechanics
analysis \cite{lopez2008}.
It would help algorithm developers and programmers
significantly if they had full confidence that
intrinsics behave correctly on branch cuts,
and no special cases need to be considered and coded for.

Expressions for these 8 complex intrinsic,
which deal correctly with $ \pm 0 , \pm \infty $
and NaN, and avoid cancellation,
were given by W. Kahan in 1987 \cite{kahan87}.
A recent study concludes that no better expressions
have been proposed since then \cite{fpa_handbook2010}.

Accuracy of complex floating point calculations was
analysed in a number of works.
Expressions for the relative
errors of complex $ \sqrt{} $ and $ \log $
(as well as $ \exp $, $ \sin $, $ \cos $)
are given in \cite{hull1994}, although the authors did
not distinguish $ + 0 $ and $ - 0 $.
The expressions are given in terms of the relative
errors of the real counterparts of these intrinsics,
e.g. their bound for the relative error in
complex $ \sqrt {} $ is $ 2 \epsilon + 1.5 E_\textrm{sqrt} $,
where $ E_\textrm{sqrt} $ is the relative error bound
for real $ \sqrt{} $.
\cite{erce2007} proposed a high speed implementation of
complex $ \sqrt{} $ which preserved the accuracy of \cite{hull1994}.
For complex $ \log $ \cite{hull1994} gives the relative
error bound of $ 3.886 \epsilon + E_{\log} $,
where $ E_{\log} $ is the relative error bound
for real $ \log x , x \gg 1 $.
For $ \arcsin $ and $ \arccos $ \cite{hull1997} give
the relative error bound of $ 9.5 \epsilon $.
The relative error bound of a fused multiply-add (FMA) for complex
multiplication was recently estimated as low as $ \epsilon $ \cite{rod2017}. 

This brief introduction shows the practical importance,
in fracture mechanics and other areas of engineering,
of such seemingly esoteric tools as signed zero and
signed infinity.
On the other hand, Fortran is still the most widely
used language for scientific and engineering computations,
particularly in high performance computing (HPC),
where Fortran codes use 60-70\% of machine cycles \cite{turner2015}.
Therefore the question of how well the above
8 complex functions are implemented in modern Fortran
deserves attention.
This question is addressed in this work
with the introduction of a set of benchmarks
with 70 tests, which
check correctness of the 8 intrinsics on branch cuts.
The code used in this work is
freely available from \url{cmplx.sf.net}.

\section{Tests}

The tests are designed to verify the behaviour
of the 8 intrinsic Fortran functions at special
points on branch cuts.
The reference expressions used for validating
the results from the test programs are derived in the Appendix.
All three IEEE
basic binary formats are verified:
binary32, binary64 and binary128 \cite{60559-2011}.

To aid portability, Fortran 2008
intrinsic module \texttt{iso\_fortran\_env}
includes named constants for these IEEE
data formats: \texttt{REAL32},
\texttt{REAL64} and \texttt{REAL128},
which are used to define the kinds of real and complex
variables and constants in the tests as e.g:

\begin{lstlisting}
use, intrinsic :: iso_fortran_env
integer, parameter :: fk=real64
real(kind=fk), parameter :: one=1.0_fk
\end{lstlisting}

The tests check that signs of the real and the imaginary
parts are correct, and
that no undue overflow, underflow or NaN
results are produced.
Fortran IEEE intrinsics
\texttt{IEEE\_CLASS},
\texttt{IEEE\_COPY\_SIGN},
\texttt{IEEE\_IS\_FINITE},
\texttt{IEEE\_IS\_NAN},
\texttt{IEEE\_SUPPORT\_SUBNORMAL},
\texttt{IEEE\_SUPPORT\_INF},
\texttt{IEEE\_SUPPORT\_NAN},
\texttt{IEEE\_VALUE}
are used, as well as
named constants
\texttt{ieee\_negative\_inf},
\texttt{ieee\_positive\_inf},
\texttt{ieee\_negative\_zero}
and
\texttt{ieee\_positive\_zero}.
For example, the values of $ \pm 0 $ and $ \pm \infty $ are
defined in the tests as:

\begin{lstlisting}
real(kind=fk) :: infp, infm, zerop, zerom
 infp=ieee_value( one, ieee_positive_inf )
 infm=ieee_value( one, ieee_negative_inf )
zerop=ieee_value( one, ieee_positive_zero )
zerom=ieee_value( one, ieee_negative_zero )
\end{lstlisting}

In addition, Fortran intrinsics
\texttt{HUGE},
\texttt{TINY}
and
\texttt{EPSILON} are used,
which return the largest and the smallest
positive model numbers respectively,
denoted $ h $ and $ t $,
and machine epsilon, $ \epsilon $.
Note that the Fortran definition of $ \epsilon $
is $ \epsilon = r^{1-p} $, where $ r $ is the radix,
$ r = 2 $ on binary computers, and $ p $ is the precision.
This definition follows the IEEE standard \cite{60559-2011}.

To help the reader understand the tests,
the maps of the branch cuts for each of the 8 functions
are given in the following sections.
The values of $ z $ and $ w $ at all tested special points
are given on each map.
The plots were calculated using \texttt{REAL64} real and complex kind
with gfortran8 compiler.

\subsection{LOG}

Behaviour of \texttt{LOG} was checked on the branch cut at 6 points:
$ z = - h \pm \ic 0 $,
$ z = - 1 \pm \ic 0 $
and
$ z = - t \pm \ic 0 $.
The top and the bottom
boundaries of the cut are mapped to
$ w = u + \ic \pi $
and
$ w = u - \ic \pi $
respectively, see Fig. \ref{fig:log1}.
 
\subsection{SQRT}

\begin{figure}
\centering
\includegraphics[width= 0.9 \textwidth]{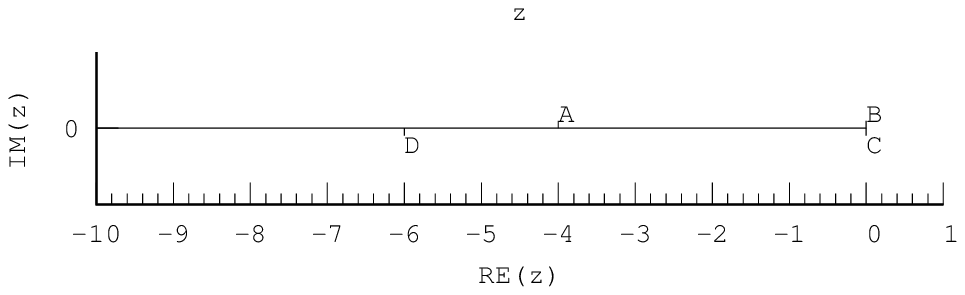} \\
\includegraphics[width= 0.9 \textwidth]{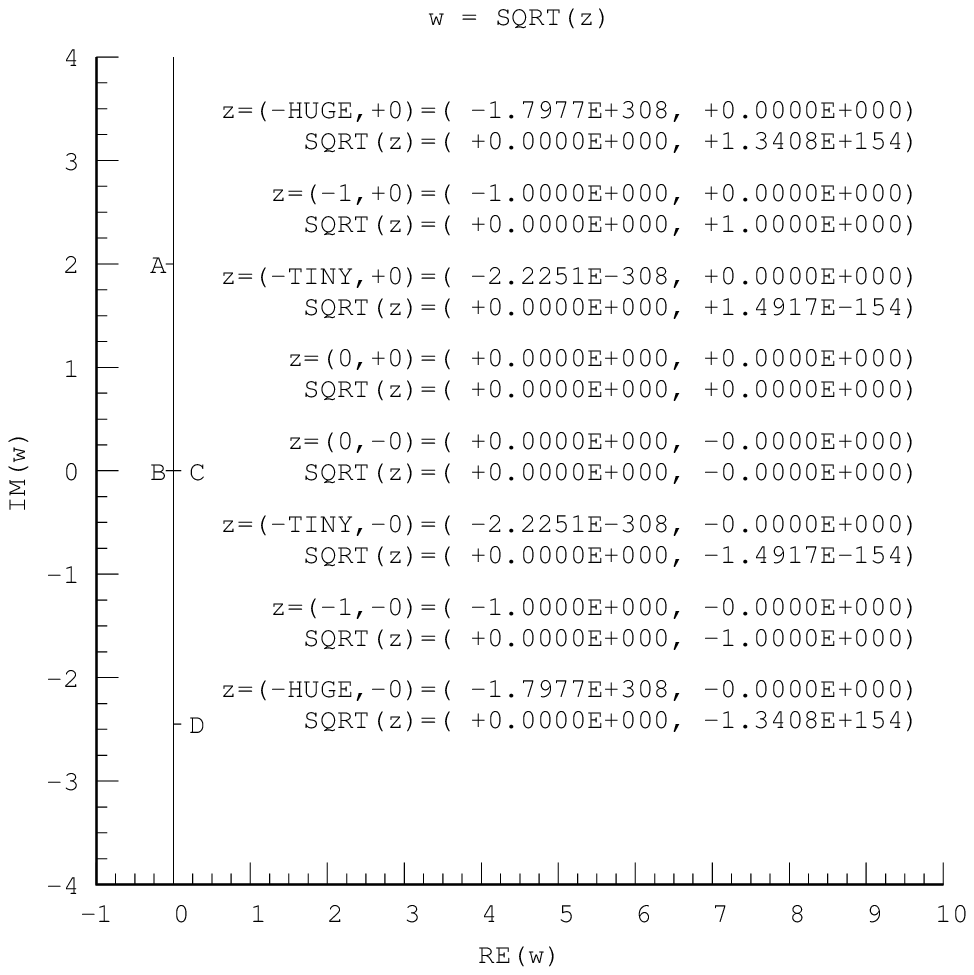}
\caption{ Map of the branch cut of $ w = \sqrt z $.
Points A and B are on the top boundary of the cut, $ y = + 0 $.
Points C and D are on the bottom boundary of the cut, $ y = - 0 $. }
\label{fig:sqrt1}
\end{figure}

Behaviour of \texttt{SQRT} was checked on the branch cut at 8 points:
$ z = - h \pm \ic 0 $,
$ z = - 1 \pm \ic 0 $,
$ z = - t \pm \ic 0 $
and
$ z = 0 \pm \ic 0 $.
The top boundary of the cut
is mapped onto the positive imaginary axis,
and the bottom boundary of the cut is mapped
onto the negative imaginary axis, see Fig. \ref{fig:sqrt1}.

\subsection{ASIN}

\begin{figure}
\centering
\includegraphics[width= 0.9 \textwidth]{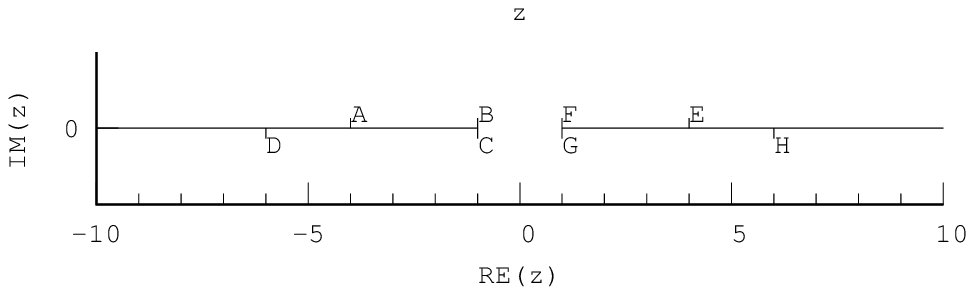} \\
\includegraphics[width= 0.9 \textwidth]{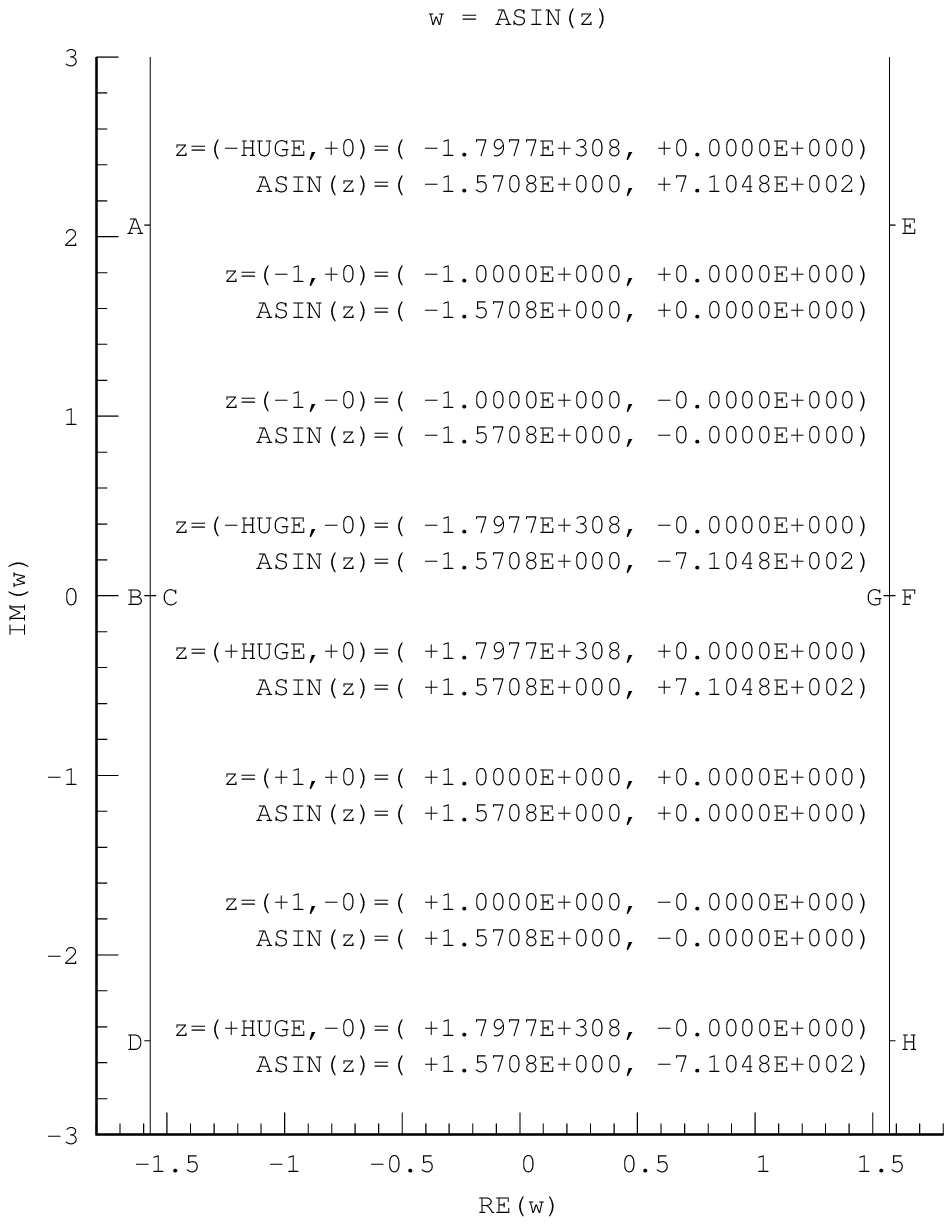}
\caption{ Map of $ w = \arcsin z $.
Points A, B, E, F are on the top boundary of the cut, $ y = + 0 $.
Points C, D, G, H are on the bottom boundary of the cut, $ y = - 0 $.
Points B and C are at $ x = -1 $.
Points F and G are at $ x = 1 $.}
\label{fig:arcsin}
\end{figure}

$ w = \arcsin z $ maps a plane with 2 cuts along the real axis,
$ x \le -1 $ and $ x \ge 1 $ to an infinite strip of width $ \pi $
along the imaginary axis,
$ - \pi  / 2 \le u \le \pi / 2 $.
The left cut, $ x \le -1 $ is mapped onto the left boundary
of the strip, $ u = - \pi / 2 $.
The right cut, $ x \ge 1 $ is mapped onto the right boundary
of the strip, $ u = \pi / 2 $, as shown in Fig. \ref{fig:arcsin}.
Behaviour of \texttt{ASIN} was checked on 8 points:
$ z = \pm h \pm \ic 0 $
and
$ z = \pm 1 \pm \ic 0 $.

\subsection{ACOS}

\begin{figure}
\centering
\includegraphics[width= 0.9 \textwidth]{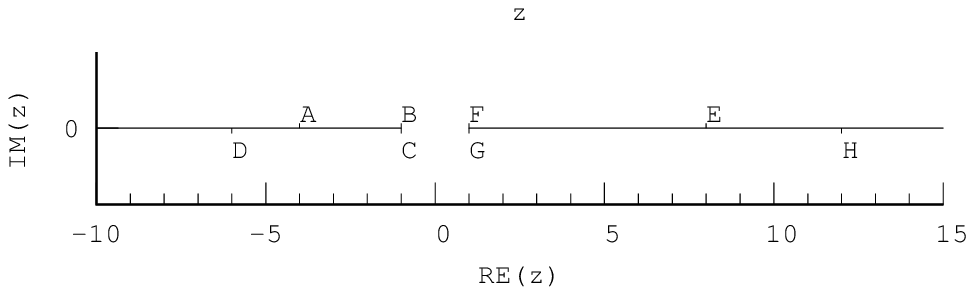} \\
\includegraphics[width= 0.9 \textwidth]{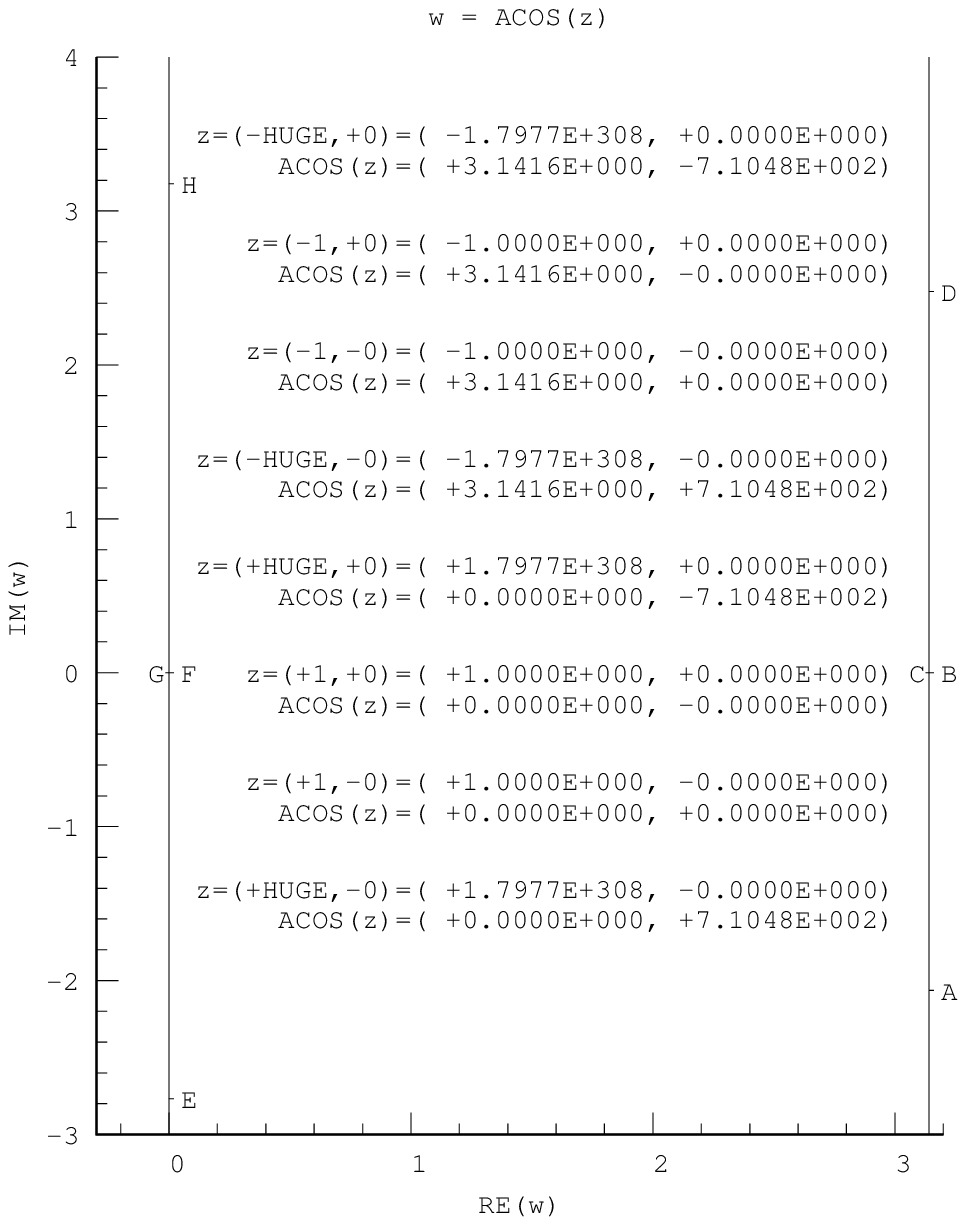}
\caption{Map of the branch cut of $ w = \arccos z $.
In plane $ z $ points A, B, E, F are on the top boundary of the cut, $ y = + 0 $.
Points C, D, G, H are on the bottom boundary of the cut, $ y = - 0 $.
Points B and C are at $ x = -1 $.
Points F and G are at $ x = 1 $.}
\label{fig:arccos}
\end{figure}

$ w = \arccos z $ has 2 branch cuts, both on the real axis,
at $ x \le - 1 $ and $ x \ge 1 $, see Fig. \ref{fig:arccos}.
For $ x \le - 1 $,
the top boundary of the cut,
$ y = + 0 $,
is mapped to
$ w = \pi - \ic b $
and the bottom boundary of the cut,
$ y = - 0 $, is mapped to
$ w = \pi + \ic b $.
For $ x \ge 1 $,
the top boundary of the cut,
$ y = + 0 $,
is mapped to
$ w = 0 - \ic b $,
and the bottom boundary of the cut,
$ y = - 0 $,
is mapped to
$ w = 0 + \ic b $.
In all cases $ b \ge 0 $.
Behaviour of \texttt{ACOS} was checked on the same 8 points
as of \texttt{ASIN}.

\subsection{ATAN}
\label{sec:atan}

\begin{figure}
\centering
\hspace*{-3mm}
\includegraphics[ height = 0.5 \textheight ]{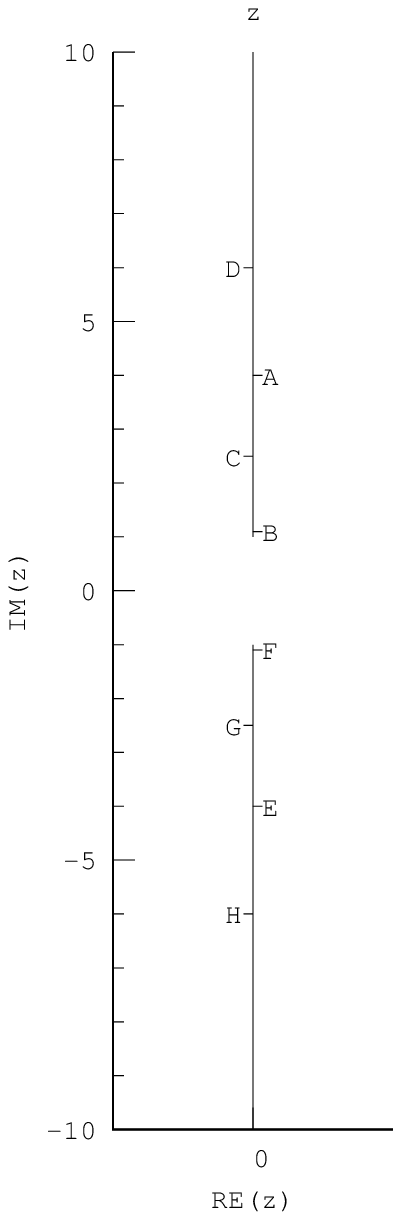} 
\includegraphics[ height = 0.5 \textheight ]{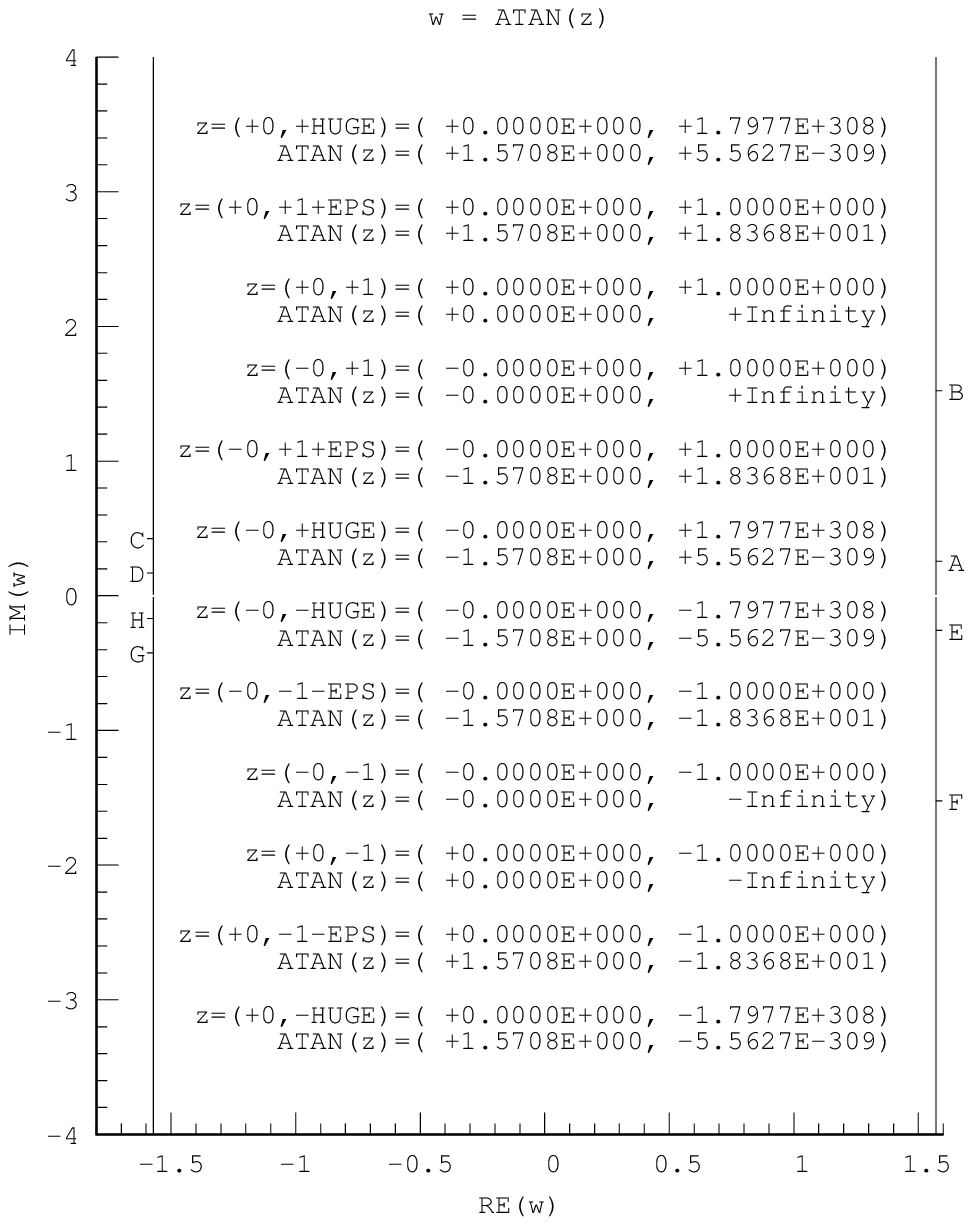}
\caption{Map of the branch cut of $ w = \arctan z $.
Points A, B, E, F are at $ x = + 0 $.
Points C, D, G, H are at $ x = - 0 $.}
\label{fig:arctan}
\end{figure}

$ w = \arctan z $ maps a plane with 2 cuts along the imaginary
axis, $ y \le - 1 $ and $ y \ge 1 $  to an infinite strip
along the imaginary axis of width $ \pi $ and centred on zero,
see Fig. \ref{fig:arctan}.

Behaviour of \texttt{ATAN} was checked in 12 points:
$ z = \pm 0 \pm \ic h $,
$ z = \pm 0 \pm \ic 1 $
and
$ z = \pm 0 \pm \ic ( 1 + \epsilon) $.
The last 4 values are interesting because they
are likely to be used
as the best substitute for $ \pm 0 \pm \ic 1 $
on systems which do not support $ \pm \infty $.

Note that in Fig. \ref{fig:arctan},
$ | \Im ( \arctan ( \pm 0 \pm \ic 1 ) ) | \approx  5.6 \times 10^{-309} $
is subnormal, the smallest \texttt{REAL64}
normal number being $ \approx 2.2 \times 10^{-308} $.
Note that \cite{60559-2011} uses the
term \emph{subnormal}
instead of the earlier \emph{denormal}.
On systems with no support for subnormals
the correct result is
$ \Im ( \arctan ( \pm 0 \pm \ic 1 ) ) = \pm 0 $,
with the correct sign.
On the other hand, on systems with no support for subnormals,
a subnormal return value is not acceptable, because
such value, $ k $, would violate the expected inequalities
$ | k | > 0 $ and $ | k | < t $, \cite{f2008}.

Also note that the exact value of
$ \Re ( \arctan( \pm 0 \pm \ic 1 ) ) $
is immaterial, provided it's finite.
For consistency with the other values of
$ \arctan $ on the cuts,
and for aesthetic pleasure,
$ \pm \pi / 2 $ might be preferred,
but the actual value has no influence on consecutive
operations made with the result of
$ \arctan ( \pm 0 \pm \ic 1 ) $,
because these will be determined
solely by the infinite imaginary part.

\subsection{ASINH}

\begin{figure}
\centering
\hspace*{-3mm}
\includegraphics[ height = 0.5 \textheight ]{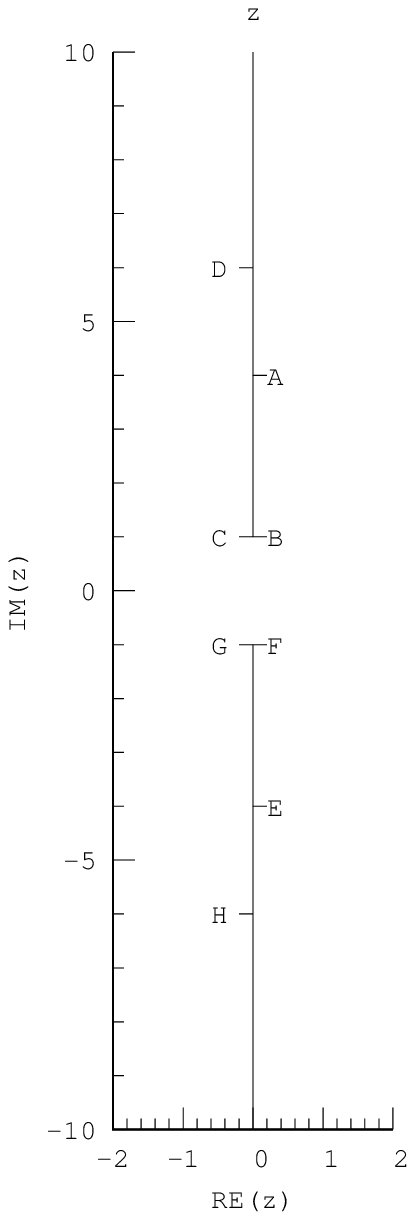} 
\includegraphics[ height = 0.5 \textheight ]{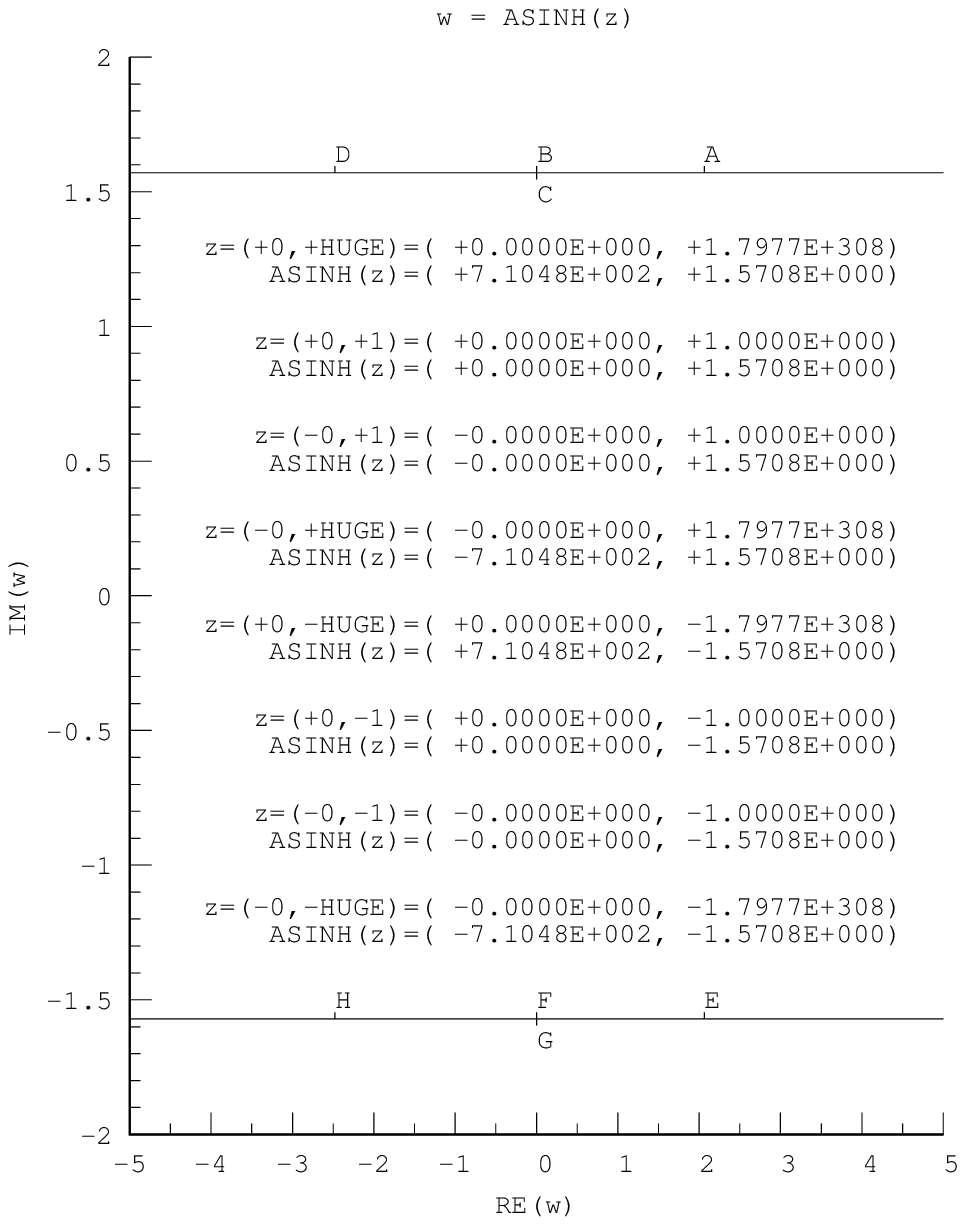}
\caption{Map of $ w = \arcsinh z $.
Points A, B, E, F are at $ x = + 0 $.
Points C, D, G, H are at $ x = - 0 $.
Points B, C are at $ y = 1 $.
Points F, G are at $ y = -1 $.}
\label{fig:arcsinh}
\end{figure}

$ w = \arcsinh z $ maps a plane with 2 cuts along the imaginary axis,
$ y \le -1 $ and $ y \ge 1 $ to an infinite strip of width $ \pi $
along the real axis,
$ - \pi / 2 \le v \le \pi / 2 $.
The bottom cut, $ y \le -1 $ is mapped onto the bottom boundary
of the strip, $ v = - \pi / 2 $.
The top cut, $ y \ge 1 $ is mapped onto the top boundary
of the strip, $ v = \pi / 2 $, as shown in Fig. \ref{fig:arcsinh}.
Behaviour of \texttt{ASINH} is checked on 8 points:
$ z = \pm 0 \pm \ic h $
and
$ z = \pm 0 \pm \ic 1 $.

\subsection{ACOSH}

\begin{figure}[ht!]
\centering
\includegraphics[ width = 0.85 \textwidth ]{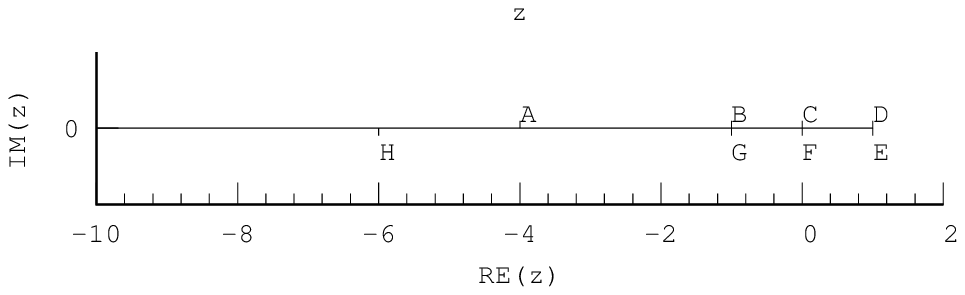} 
\includegraphics[ width = 0.85 \textwidth ]{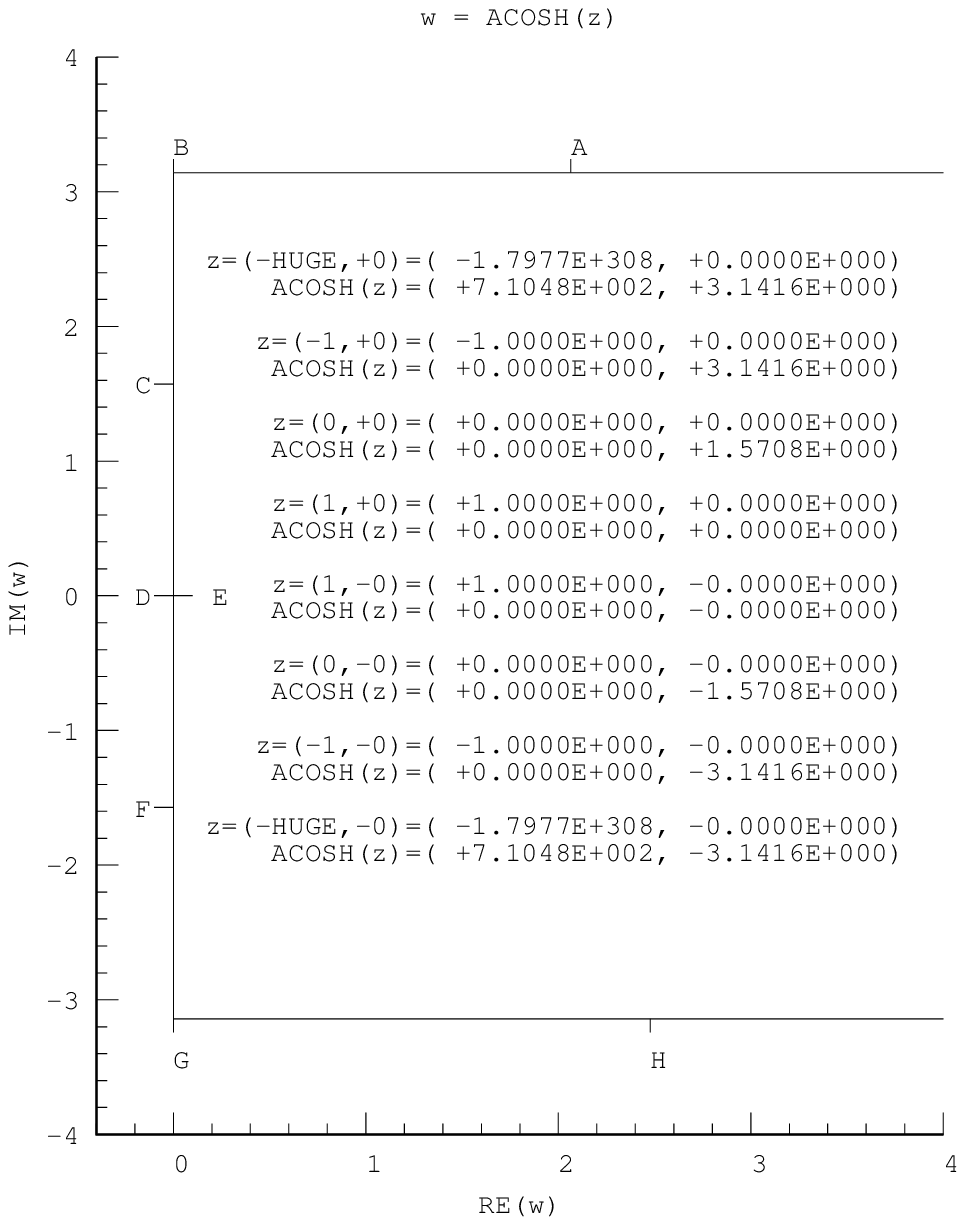}
\caption{Map of $ w = \arccosh z $.
Points A, B, E, F are at $ x = + 0 $.
Points C, D, G, H are at $ x = - 0 $.
Points B, C are at $ y = 1 $.
Points F, G are at $ y = -1 $.}
\label{fig:arccosh}
\end{figure}

$ w = \arccosh z $ maps a plane with a single cut
along the real axis at $ x \le 1 $ onto a semi-infinite
strip of width $ 2 \pi $, running along the real axis,
$ u \ge 0 $, see Fig. \ref{fig:arccosh}.

The tests check that
(1) the top side of the cut at $ x \le -1 $ is mapped
onto the top boundary of the strip, $ u \ge 0 , v = \pi $;
(2) the top side of the cut at $ -1 \le x \le 1 $
is mapped onto the end of the strip at $ u = 0 , 0 \le v \le \pi $;
(3) the bottom side of the cut at $ -1 \le x \le 1 $
is mapped onto the end of the strip at $ u = 0 , - \pi \le v \le 0 $,
and
(4) the bottom side of the cut at $ x \le -1 $ 
is mapped onto the
bottom boundary of the strip, $ u \ge 0 , v = - \pi $.
Thus, behaviour of \texttt{ACOSH} is checked on 8 points:
$ z = -h \pm \ic 0 $,
$ z = -1 \pm \ic 0 $,
$ z = 0 \pm \ic 0 $
and
$ z = 1 \pm \ic 0 $.

\subsection{ATANH}

\begin{figure}
\centering
\includegraphics[ width = 0.9 \textwidth ]{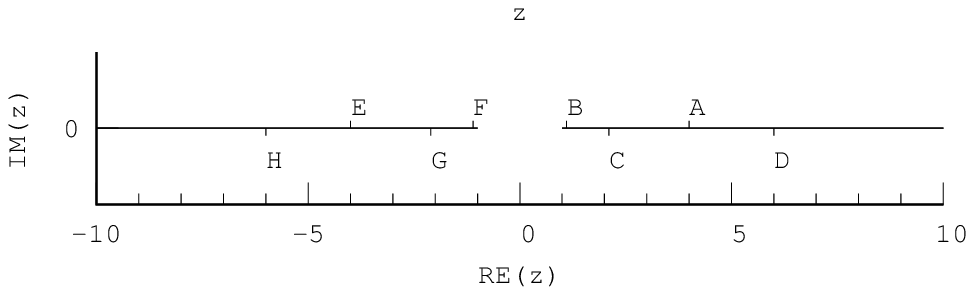}  \\
\includegraphics[ width = 0.9 \textwidth ]{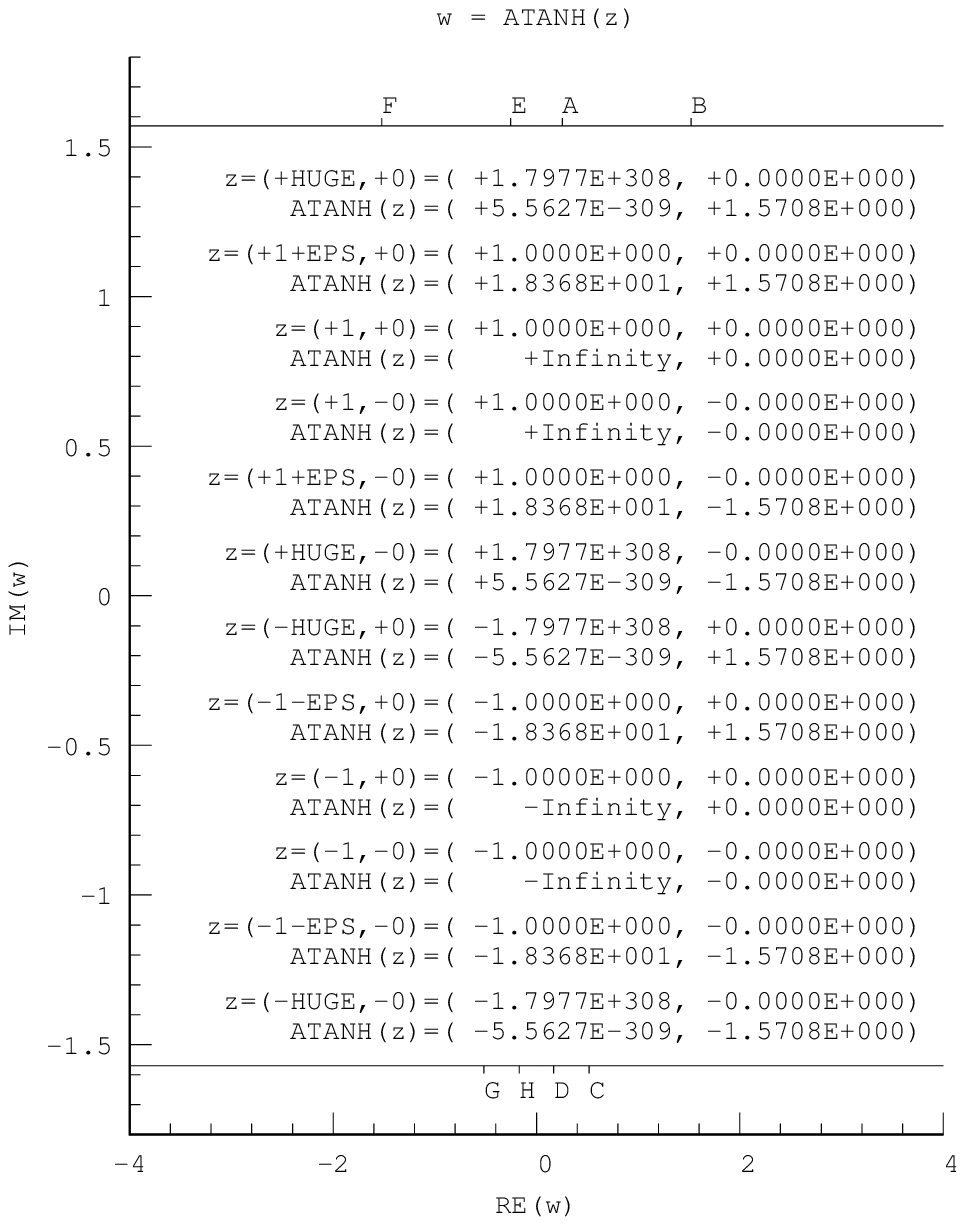}
\caption{Map of $ w = \arctanh z $.
Points A, B, E, F are at $ x = + 0 $.
Points C, D, G, H are at $ x = - 0 $.
Points B and C are at $ y = 1 $.
Points F and G are at $ y = -1 $.}
\label{fig:arctanh}
\end{figure}

$ w = \arctanh z $ maps a plane with 2 cuts
along the real axis, $ x \le -1 $ and $ x \ge 1 $
onto a infinite strip of width $ \pi $ centered on
0 and running along the real axis, see Fig. \ref{fig:arctanh}.
\texttt{ATANH} was verified on 12 points:
$ z = \pm h \pm \ic 0 $,
$ z = \pm 1 \pm \ic 0 $
and
$ z = \pm ( 1 + \epsilon ) \pm \ic 0 $.
Behaviour of \texttt{ATANH} on the branch cuts
mirrors many features of that of \texttt{ATAN},
see Sec. \ref{sec:atan}.

\section{Results}
\label{sec:res}

Fortran compilers, compiler options
operating systems and CPUs used in
this study are detailed in Tab. \ref{tab:compilers}.

\begin{table}[h]
\centering
\begin{small}
\begin{tabular} { l l l l }
\hline
Compiler & Compiler options & OS & CPU \\
\hline
gfortran 8.0 & \texttt{-fsign-zero} & FreeBSD & Haswell \\
Flang 4.0    & \texttt{-Kieee} & FreeBSD & Haswell \\
Cray 8.5.8   & \texttt{-hfp0} & linux & IvyBridge \\
Oracle 12.6 Fortran 95 8.8 & \texttt{-fsimple=0 -ftrap=none} & linux & SandyBridge \\
PGI 16.3-0   & \texttt{-Kieee} & linux & SandyBridge \\
Intel 18.0.1 & \texttt{-assume minus0} & linux & Broadwell \\ 
NAG 6.2(Chiyoda) Build 6201 & \texttt{-ieee=full} & linux & Broadwell \\
IBM XL Fortran 15.1.5 & \texttt{-qstrict} & linux & Power8 \\
\hline
\end{tabular}
\end{small}
\caption{Compilers, compiler options, operating
systems and CPUs used in this work.}
\label{tab:compilers}
\end{table}

Flang (\texttt{github.com/flang-compiler})
is an open source front end
based on Nvidia PGI compiler, targeting LLVM.
Flang 4.0 does not support inverse hyperbolic intrinsics. 

Oracle 12.6 Fortran 95 compiler, released in May 2017,
supports some Fortran 2003 and 2008 features, but
not complex arguments for inverse trigonometric intrinsics.

PGI 16.3-0 does not support complex arguments to
inverse hyperbolic intrinsics.

The results are presented in the form of tables.
The dot entry, "\ok", means the test has passed.
Several kinds of test failures are distinguished,
as detailed in Tab. \ref{tab:entries}.

\begin{table}[h]
\centering
\begin{tabular} { c l }
\hline
Symbol & Failure type \\
\hline
$ \times $ & intrinsic not implemented with this argument \\
$ d $ & subnormal value returned, but subnormals are not supported \\
$ m $ & wrong magnitude of finite non-zero real/imaginary part \\
$ n $ & NaN, the correct value is finite or infinite \\
$ o $ & unwarranted overflow, the correct value is finite \\
$ p $ & normal finite non-zero result, the correct value is 0 \\
$ s $ & wrong sign of real/imaginary part, or both \\
$ z $ & zero real/imaginary part, the correct value is normal finite non-zero \\
\hline
\end{tabular}
\caption{List of test failure types.
The symbols are used in Tabs. \ref{tab:monkey},
\ref{tab:apple} and \ref{tab:pear}. }
\label{tab:entries}
\end{table}

Multiple failures are possible in a single test,
e.g. an entry "$ osz $"
means that failures of kind "$ o $", "$ s $" and
"$ z $" have occurred in that test.
Most failure types are self-explanatory, except type "$ m $",
which is justified in App. \ref{sec:log2h}.

\subsection{ \texttt{REAL32} }

For \texttt{REAL32} real and complex variables
$ h \approx 3.4 \times 10^{38} $,
$ t \approx 1.2  \times 10^{-38} $
and
$ \epsilon \approx 1.2 \times 10^{-7} $.
The test results are summarised in Tab. \ref{tab:monkey}.

\begin{footnotesize}
\begin{longtable} { l c c c c c c c c }
%\begin{equation*}
%\begin{array}{ l c c c c c c c c }
Test & GCC & Flang & Cray & Oracle & PGI & Intel & NAG & IBM \\
\hline
\endhead
\hline
\\
\caption{\normalsize Test results for \texttt{REAL32} kind.
$ t = $\texttt{ TINY(0.0\_REAL32)}; $ h = $\texttt{ HUGE(0.0\_REAL32)};
$ e = $\texttt{ EPSILON(0.0\_REAL32)}.}
\endfoot
$ \log( -h + \ic 0 ) $ &
 \ok & \ok & \ok & \ok & \ok & \ok & \ok & \ok \\
$ \log( -1 + \ic 0 ) $ &
 \ok & \ok & \ok & \ok & \ok & \ok & \ok & \ok \\
$ \log( -t + \ic 0 ) $ &
 \ok & \ok & \ok & \ok & \ok & \ok & \ok & \ok \\
$ \log( -t - \ic 0 ) $ & 
 \ok & \ok & \ok & \ok & \ok & \ok & \ok & \ok \\
$ \log( -1 - \ic 0 ) $ &
 \ok & \ok & \ok & \ok & \ok & \ok & \ok & \ok \\
$ \log(-h - \ic 0 ) $ &
 \ok & \ok & \ok & \ok & \ok & \ok & \ok & \ok \\
\hline
$ \sqrt{ -h + \ic 0 } $ &
 \ok & $ o $ & \ok & \ok   & $ o $  & \ok & \ok & \ok \\
$ \sqrt{ -1 + \ic 0 } $ &
 \ok & \ok   & \ok & \ok   & \ok    & \ok & \ok & \ok \\
$ \sqrt{ -t + \ic 0 } $ &
 \ok & \ok   & \ok & \ok   & \ok    & \ok & \ok & \ok \\
$ \sqrt{ 0  + \ic 0 } $ &
 \ok & \ok   & \ok & $ s $ & \ok    & \ok & \ok & \ok \\
$ \sqrt{ 0  - \ic 0 } $ &
 \ok & $ s $ & \ok & $ s $ & $ s $  & \ok & \ok & $ s $ \\
$ \sqrt{ -t - \ic 0 } $ &
 \ok & \ok   & \ok & \ok   & $ s $  & \ok & \ok & \ok \\
$ \sqrt{ -1 - \ic 0 } $ &
 \ok & \ok   & \ok & \ok   & $ s $  & \ok & \ok & \ok \\
$ \sqrt{ -h - \ic 0 } $ &
 \ok & $ o $ & \ok & \ok   & $ os $ & \ok & \ok & \ok \\
\hline
$ \arcsin( - h + \ic 0 ) $ &
 \ok & \ok & $ os $ & $ \times $ & $ os $ & \ok & \ok & $ o $ \\
$ \arcsin( - 1 + \ic 0 ) $ &
 \ok & \ok & $ s  $ & $ \times $ & $ s $  & \ok & \ok & \ok \\
$ \arcsin( - 1 - \ic 0 ) $ &
 \ok & \ok & \ok    & $ \times $ & \ok    & \ok & \ok & \ok \\
$ \arcsin( - h - \ic 0 ) $ &
 \ok & \ok & $ o  $ & $ \times $ & $ os $ & \ok & \ok & $ o $ \\
$ \arcsin(   h + \ic 0 ) $ &
 \ok & \ok & $ os $ & $ \times $ & $ os $ & \ok & \ok & $ o $ \\
$ \arcsin(   1 + \ic 0 ) $ &
 \ok & \ok & $ s  $ & $ \times $ & $ s $  & \ok & \ok & \ok \\
$ \arcsin(   1 - \ic 0 ) $ &
 \ok & \ok & \ok    & $ \times $ & \ok    & \ok & \ok & \ok \\
$ \arcsin(   h - \ic 0 ) $ &
 \ok & \ok & $ o $  & $ \times $ & $ os $ & \ok & \ok & $ o $ \\
\hline
$ \arccos( - h + \ic 0 ) $ &
 \ok & \ok & $ osz $ & $ \times $ & $ osz $ & \ok & \ok & $ o $ \\
$ \arccos( - 1 + \ic 0 ) $ &
 \ok & \ok & $ s   $ & $ \times $ & $ s $   & \ok & \ok & \ok \\
$ \arccos( - 1 - \ic 0 ) $ &
 \ok & \ok & \ok     & $ \times $ & \ok     & \ok & \ok & \ok \\
$ \arccos( - h - \ic 0 ) $ &
 \ok & \ok & $ o   $ & $ \times $ & $ oz $  & \ok & \ok & $ o $ \\
$ \arccos(   h + \ic 0 ) $ &
 \ok & \ok & $ osz $ & $ \times $ & $ ops $ & \ok & \ok & $ o $ \\
$ \arccos(   1 + \ic 0 ) $ &
 \ok & \ok & $ s   $ & $ \times $ & $ s $   & \ok & \ok & \ok \\
$ \arccos(   1 - \ic 0 ) $ &
 \ok & \ok & \ok     & $ \times $ & \ok     & \ok & \ok & \ok \\
$ \arccos(   h - \ic 0 ) $ &
 \ok & \ok & $ o   $ & $ \times $ & $ op $  & \ok & \ok & $ o $ \\
\hline
$ \arctan( + 0 + \ic h ) $ &
 \ok & \ok   & $ n $ & $ \times $ & $ n $  & \ok & $ nz $ & $ n $ \\
$ \arctan( + 0 + \ic ( 1 + e ) ) $ &
 \ok & \ok   & \ok   & $ \times $ & \ok    & \ok & $ nz $ & \ok \\
$ \arctan( + 0 + \ic 1 ) $ &
 \ok & \ok   & \ok   & $ \times $ & \ok    & \ok & \ok    & \ok \\
$ \arctan( - 0 + \ic 1 ) $ &
 \ok & \ok   & \ok   & $ \times $ & \ok    & \ok & \ok    & \ok \\
$ \arctan( - 0 + \ic ( 1 + e ) ) $ &
 \ok & $ s $ & \ok   & $ \times $ & $ s $  & \ok & $ nz $ & \ok \\
$ \arctan( - 0 + \ic h ) $ &
 \ok & $ s $ & $ n $ & $ \times $ & $ ns $ & \ok & $ nz $ & $ n $ \\
$ \arctan( - 0 - \ic h ) $ &
 \ok & \ok   & $ n $ & $ \times $ & $ n $  & \ok & $ nz $ & $ n $ \\
$ \arctan( - 0 - \ic ( 1 + e ) ) $ &
 \ok & \ok   & \ok   & $ \times $ & \ok    & \ok & $ nz $ & \ok \\
$ \arctan( - 0 - \ic 1 ) $ &
 \ok & \ok   & \ok   & $ \times $ & \ok    & \ok & \ok    & \ok \\
$ \arctan( + 0 - \ic 1 ) $ &
 \ok & \ok   & \ok   & $ \times $ & \ok    & \ok & \ok    & \ok \\
$ \arctan( + 0 - \ic ( 1 + e ) ) $ &
 \ok & \ok   & \ok   & $ \times $ & \ok    & \ok & $ nz $ & \ok \\
$ \arctan( + 0 - \ic h ) $ &
 \ok & \ok   & $ n $ & $ \times $ & $ n $  & \ok & $ nz $ & $ n $ \\
\hline
$ \arcsinh ( + 0 + \ic h ) $ &
 \ok & $ \times $ & $ o $  & $ o $  & $ \times $ & \ok & \ok & $ o $ \\
$ \arcsinh ( + 0 + \ic 1 ) $ &
 \ok & $ \times $ & \ok    & \ok    & $ \times $ & \ok & \ok & \ok \\
$ \arcsinh ( - 0 + \ic 1 ) $ &
 \ok & $ \times $ & $ s $  & $ s $  & $ \times $ & \ok & \ok & \ok \\
$ \arcsinh ( - 0 + \ic h ) $ &
 \ok & $ \times $ & $ os $ & $ os $ & $ \times $ & \ok & \ok & $ o $ \\
$ \arcsinh ( + 0 - \ic h ) $ &
 \ok & $ \times $ & $ o $  & $ o $  & $ \times $ & \ok & \ok & $ o $ \\
$ \arcsinh ( + 0 - \ic 1 ) $ &
 \ok & $ \times $ & \ok    & \ok    & $ \times $ & \ok & \ok & \ok \\
$ \arcsinh ( - 0 - \ic 1 ) $ &
 \ok & $ \times $ & $ s $  & $ s $  & $ \times $ & \ok & \ok & \ok \\
$ \arcsinh ( - 0 - \ic h ) $ &
 \ok & $ \times $ & $ os $ & $ os $ & $ \times $ & \ok & \ok & $ o $ \\
\hline
$ \arccosh ( + 0 + \ic h ) $ &
 \ok & $ \times $ & $ o $ & $ o $ & $ \times $ & \ok & $ m $  & \ok \\
$ \arccosh ( + 0 + \ic 1 ) $ &
 \ok & $ \times $ & \ok   & \ok   & $ \times $ & \ok & \ok    & \ok \\
$ \arccosh ( - 0 + \ic 1 ) $ &
 \ok & $ \times $ & \ok   & \ok   & $ \times $ & \ok & $ p $  & \ok \\
$ \arccosh ( - 0 + \ic h ) $ &
 \ok & $ \times $ & \ok   & \ok   & $ \times $ & \ok & \ok    & \ok \\
$ \arccosh ( + 0 - \ic h ) $ &
 \ok & $ \times $ & \ok   & \ok   & $ \times $ & \ok & $ s $  & \ok \\
$ \arccosh ( + 0 - \ic 1 ) $ &
 \ok & $ \times $ & \ok   & \ok   & $ \times $ & \ok & $ ps $ & \ok \\
$ \arccosh ( - 0 - \ic 1 ) $ &
 \ok & $ \times $ & \ok   & \ok   & $ \times $ & \ok & $ s $  & \ok \\
$ \arccosh ( - 0 - \ic h ) $ &
 \ok & $ \times $ & $ o $ & $ o $ & $ \times $ & \ok & $ ms $ & \ok \\
\hline
$ \arctanh ( h     + \ic 0 ) $ &
 \ok & $ \times $ & $ n $ & $ n $ & $ \times $ & \ok & $ nz $ & $ o $ \\
$ \arctanh ( 1 + e + \ic 0 ) $ &
 \ok & $ \times $ & \ok   & \ok   & $ \times $ & \ok & $ nz $ & \ok \\
$ \arctanh ( 1     + \ic 0 ) $ &
 \ok & $ \times $ & \ok   & \ok   & $ \times $ & \ok & \ok    & \ok \\
$ \arctanh ( 1     - \ic 0 ) $ &
 \ok & $ \times $ & \ok   & \ok   & $ \times $ & \ok & \ok    & \ok \\
$ \arctanh ( 1 + e - \ic 0 ) $ &
 \ok & $ \times $ & \ok   & \ok   & $ \times $ & \ok & $ nz $ & \ok \\
$ \arctanh ( h     - \ic 0 ) $ &
 \ok & $ \times $ & $ n $ & $ n $ & $ \times $ & \ok & $ nz $ & $ o $ \\
$ \arctanh ( -h    + \ic 0 ) $ &
 \ok & $ \times $ & $ n $ & $ n $ & $ \times $ & \ok & $ nz $ & $ o $ \\
$ \arctanh ( -1 -e + \ic 0 ) $ &
 \ok & $ \times $ & \ok   & \ok   & $ \times $ & \ok & $ nz $ & \ok \\
$ \arctanh ( - 1   + \ic 0 ) $ &
 \ok & $ \times $ & \ok   & \ok   & $ \times $ & \ok & \ok    & \ok \\
$ \arctanh ( - 1   - \ic 0 ) $ &
 \ok & $ \times $ & \ok   & \ok   & $ \times $ & \ok & \ok    & \ok \\
$ \arctanh ( -1 -e - \ic 0 ) $ &
 \ok & $ \times $ & \ok   & \ok   & $ \times $ & \ok & $ nz $ & \ok \\
$ \arctanh ( - h   - \ic 0 ) $ &
 \ok & $ \times $ & $ n $ & $ n $ & $ \times $ & \ok & $ nz $ & $ o $ \\
\hline
Pass rate & 70/70 & 37/42 & 42/70 & 28/42 & 20/42 & 70/70 & 48/70 & 49/70
%\end{array}
%\end{equation*}
\label{tab:monkey}
\end{longtable}
\end{footnotesize}

\subsection{ \texttt{REAL64} }

For \texttt{REAL64} real and complex variables
$ h \approx 1.8 \times 10^{308} $,
$ t \approx 2.2  \times 10^{-308} $
and
$ \epsilon \approx 2.2 \times 10^{-16} $.
The test results are summarised in Tab. \ref{tab:apple}.

\begin{footnotesize}
\begin{longtable} { l c c c c c c c c }
%\begin{equation*}
%\begin{array}{ l c c c c c c c c }
Test & GCC & Flang & Cray & Oracle & PGI & Intel & NAG & IBM \\
\hline
\endhead
\hline
\\
\caption{\normalsize Test results for \texttt{REAL64} kind.
$ t = $\texttt{ TINY(0.0\_REAL64)}; $ h = $\texttt{ HUGE(0.0\_REAL64)};
$ e = $\texttt{ EPSILON(0.0\_REAL64)}.}
\endfoot
$ \log( -h + \ic 0 ) $ &
 \ok & \ok & \ok & \ok & \ok & \ok & \ok & \ok \\
$ \log( -1 + \ic 0 ) $ &
 \ok & \ok & \ok & \ok & \ok & \ok & \ok & \ok \\
$ \log( -t + \ic 0 ) $ &
 \ok & \ok & \ok & \ok & \ok & \ok & \ok & \ok \\
$ \log( -t - \ic 0 ) $ &
 \ok & \ok & \ok & \ok & \ok & \ok & \ok & \ok \\
$ \log( -1 - \ic 0 ) $ &
 \ok & \ok & \ok & \ok & \ok & \ok & \ok & \ok \\
$ \log( -h - \ic 0 ) $ &
 \ok & \ok & \ok & \ok & \ok & \ok & \ok & \ok \\
\hline
$ \sqrt{ -h + \ic 0 } $ &
 \ok & $ o $ & \ok   & $ o $ & $ o $  & \ok & \ok & $ o $ \\
$ \sqrt{ -1 + \ic 0 } $ &
 \ok & \ok   & \ok   & \ok   & \ok    & \ok & \ok & \ok \\
$ \sqrt{ -t + \ic 0 } $ &
 \ok & \ok   & \ok   & \ok   & \ok    & \ok & \ok & \ok \\
$ \sqrt{ 0  + \ic 0 } $ &
 \ok & \ok   & \ok   & \ok   & \ok    & \ok & \ok & \ok \\
$ \sqrt{ 0  - \ic 0 } $ &
 \ok & $ s $ & \ok   & \ok   & $ s $  & \ok & \ok & $ s $ \\
$ \sqrt{ -t - \ic 0 } $ &
 \ok & \ok   & \ok   & \ok   & $ s $  & \ok & \ok & \ok \\
$ \sqrt{ -1 - \ic 0 } $ &
 \ok & \ok   & \ok   & \ok   & $ s $  & \ok & \ok & \ok \\
$ \sqrt{ -h - \ic 0 } $ &
 \ok & $ o $ & $ o $ & $ o $ & $ os $ & \ok & \ok & $ o $ \\
\hline
$ \arcsin( - h + \ic 0 ) $ &
 \ok & \ok & $ os $ & $ \times $ & $ os $ & \ok & \ok & $ o $ \\
$ \arcsin( - 1 + \ic 0 ) $ &
 \ok & \ok & $ s $  & $ \times $ & $ s $  & \ok & \ok & \ok \\
$ \arcsin( - 1 - \ic 0 ) $ &
 \ok & \ok & \ok    & $ \times $ & \ok    & \ok & \ok & \ok \\
$ \arcsin( - h - \ic 0 ) $ &
 \ok & \ok & $ o $  & $ \times $ & $ os $ & \ok & \ok & $ o $ \\
$ \arcsin(   h + \ic 0 ) $ &
 \ok & \ok & $ os $ & $ \times $ & $ os $ & \ok & \ok & $ o $ \\
$ \arcsin(   1 + \ic 0 ) $ &
 \ok & \ok & $ s $  & $ \times $ & $ s $  & \ok & \ok & \ok \\
$ \arcsin(   1 - \ic 0 ) $ &
 \ok & \ok & \ok    & $ \times $ & \ok    & \ok & \ok & \ok \\
$ \arcsin(   h - \ic 0 ) $ &
 \ok & \ok & $ o $  & $ \times $ & $ os $ & \ok & \ok & $ o $ \\
\hline
$ \arccos( - h + \ic 0 ) $ &
 \ok & \ok & $ osz $ & $ \times $ & $ osz $ & \ok & \ok & $ o $ \\
$ \arccos( - 1 + \ic 0 ) $ &
 \ok & \ok & $ s $   & $ \times $ & $ s $   & \ok & \ok & \ok \\
$ \arccos( - 1 - \ic 0 ) $ &
 \ok & \ok & \ok     & $ \times $ & \ok     & \ok & \ok & \ok \\
$ \arccos( - h - \ic 0 ) $ &
 \ok & \ok & $ o $   & $ \times $ & $ oz $  & \ok & \ok & $ o $ \\
$ \arccos(   h + \ic 0 ) $ &
 \ok & \ok & $ ops $ & $ \times $ & $ ops $ & \ok & \ok & $ o $ \\
$ \arccos(   1 + \ic 0 ) $ &
 \ok & \ok & $ s $   & $ \times $ & $ s $   & \ok & \ok & \ok \\
$ \arccos(   1 - \ic 0 ) $ &
 \ok & \ok & \ok     & $ \times $ & \ok     & \ok & \ok & \ok \\
$ \arccos(   h - \ic 0 ) $ &
 \ok & \ok & $ o $   & $ \times $ & $ op $  & \ok & \ok & $ o $ \\
\hline
$ \arctan( + 0 + \ic h ) $ &
 \ok & \ok   & $ n $ & $ \times $ & $ n $  & $ d $ & $ nz $ & $ n $ \\
$ \arctan( + 0 + \ic ( 1 + e ) ) $ &
 \ok & \ok   & \ok   & $ \times $ & \ok    & \ok   & $ nz $ & \ok \\
$ \arctan( + 0 + \ic 1 ) $ &
 \ok & \ok   & \ok   & $ \times $ & \ok    & \ok   & \ok    & \ok \\
$ \arctan( - 0 + \ic 1 ) $ &
 \ok & \ok   & \ok   & $ \times $ & \ok    & \ok   & \ok    & \ok \\
$ \arctan( - 0 + \ic ( 1 + e ) ) $ &
 \ok & $ s $ & \ok   & $ \times $ & $ s $  & \ok   & $ nz $ & \ok \\
$ \arctan( - 0 + \ic h ) $ &
 \ok & $ s $ & $ n $ & $ \times $ & $ ns $ & $ d $ & $ nz $ & $ n $ \\
$ \arctan( - 0 - \ic h ) $ &
 \ok & \ok   & $ n $ & $ \times $ & $ n $  & $ d $ & $ nz $ & $ n $ \\
$ \arctan( - 0 - \ic ( 1 + e ) ) $ &
 \ok & \ok   & \ok   & $ \times $ & \ok    & \ok   & $ nz $ & \ok \\
$ \arctan( - 0 - \ic 1 ) $ &
 \ok & \ok   & \ok   & $ \times $ & \ok    & \ok   & \ok    & \ok \\
$ \arctan( + 0 - \ic 1 ) $ &
 \ok & \ok   & \ok   & $ \times $ & \ok    & \ok   & \ok    & \ok \\
$ \arctan( + 0 - \ic ( 1 + e ) ) $ &
 \ok & \ok   & \ok   & $ \times $ & \ok    & \ok   & $ nz $ & \ok \\
$ \arctan( + 0 - \ic h ) $ &
 \ok & \ok   & $ n $ & $ \times $ & $ n $  & $ d $ & $ nz $ & $ n $ \\
\hline
$ \arcsinh ( + 0 + \ic h ) $ &
 \ok & $ \times $ & $ o $  & $ o $  & $ \times $ & \ok & \ok & $ o $ \\
$ \arcsinh ( + 0 + \ic 1 ) $ &
 \ok & $ \times $ & \ok    & \ok    & $ \times $ & \ok & \ok & \ok \\
$ \arcsinh ( - 0 + \ic 1 ) $ &
 \ok & $ \times $ & $ s $  & $ s $  & $ \times $ & \ok & \ok & \ok \\
$ \arcsinh ( - 0 + \ic h ) $ &
 \ok & $ \times $ & $ os $ & $ os $ & $ \times $ & \ok & \ok & $ o $ \\
$ \arcsinh ( + 0 - \ic h ) $ &
 \ok & $ \times $ & $ o $  & $ o $  & $ \times $ & \ok & \ok & $ o $ \\
$ \arcsinh ( + 0 - \ic 1 ) $ &
 \ok & $ \times $ & \ok    & \ok    & $ \times $ & \ok & \ok & \ok \\
$ \arcsinh ( - 0 - \ic 1 ) $ &
 \ok & $ \times $ & $ s $  & $ s $  & $ \times $ & \ok & \ok & \ok \\
$ \arcsinh ( - 0 - \ic h ) $ &
 \ok & $ \times $ & $ os $ & $ os $ & $ \times $ & \ok & \ok & $ o $ \\
\hline
$ \arccosh ( + 0 + \ic h ) $ &
 \ok & $ \times $ & $ no $ & $ n $ & $ \times $ & \ok & $ no $ & \ok \\
$ \arccosh ( + 0 + \ic 1 ) $ &
 \ok & $ \times $ & \ok    & \ok   & $ \times $ & \ok & \ok    & \ok \\
$ \arccosh ( - 0 + \ic 1 ) $ &
 \ok & $ \times $ & \ok    & \ok   & $ \times $ & \ok & \ok    & \ok\\
$ \arccosh ( - 0 + \ic h ) $ &
 \ok & $ \times $ & \ok    & \ok   & $ \times $ & \ok & \ok    & \ok \\
$ \arccosh ( + 0 - \ic h ) $ &
 \ok & $ \times $ & \ok    & \ok   & $ \times $ & \ok & $ s $  & \ok \\
$ \arccosh ( + 0 - \ic 1 ) $ &
 \ok & $ \times $ & \ok    & \ok   & $ \times $ & \ok & $ s $  & \ok \\
$ \arccosh ( - 0 - \ic 1 ) $ &
 \ok & $ \times $ & \ok    & \ok   & $ \times $ & \ok & $ s $  & \ok \\
$ \arccosh ( - 0 - \ic h ) $ &
 \ok & $ \times $ & $ no $ & $ n $ & $ \times $ & \ok & $ no $ & \ok \\
\hline
$ \arctanh ( h     + \ic 0 ) $ &
 \ok & $ \times $ & $ n $ & $ n $ & $ \times $ & $ d $ & $ nz $ & $ n $ \\
$ \arctanh ( 1 + e + \ic 0 ) $ &
 \ok & $ \times $ & \ok   & \ok   & $ \times $ & \ok   & $ nz $ & \ok \\
$ \arctanh ( 1     + \ic 0 ) $ &
 \ok & $ \times $ & \ok   & \ok   & $ \times $ & \ok   & \ok    & \ok \\
$ \arctanh ( 1     - \ic 0 ) $ &
 \ok & $ \times $ & \ok   & \ok   & $ \times $ & \ok   & \ok    & \ok \\
$ \arctanh ( 1 + e - \ic 0 ) $ &
 \ok & $ \times $ & \ok   & \ok   & $ \times $ & \ok   & $ nz $ & \ok \\
$ \arctanh ( h     - \ic 0 ) $ &
 \ok & $ \times $ & $ n $ & $ n $ & $ \times $ & $ d $ & $ nz $ & $ n $ \\
$ \arctanh ( -h    + \ic 0 ) $ &
 \ok & $ \times $ & $ n $ & $ n $ & $ \times $ & $ d $ & $ nz $ & $ n $ \\
$ \arctanh ( -1 -e + \ic 0 ) $ &
 \ok & $ \times $ & \ok   & \ok   & $ \times $ & \ok   & $ nz $ & \ok \\
$ \arctanh ( - 1   + \ic 0 ) $ &
 \ok & $ \times $ & \ok   & \ok   & $ \times $ & \ok   & \ok    & \ok \\
$ \arctanh ( - 1   - \ic 0 ) $ &
 \ok & $ \times $ & \ok   & \ok   & $ \times $ & \ok   & \ok    & \ok \\
$ \arctanh ( -1 -e - \ic 0 ) $ &
 \ok & $ \times $ & \ok   & \ok   & $ \times $ & \ok   & $ nz $ & \ok \\
$ \arctanh ( - h   - \ic 0 ) $ &
 \ok & $ \times $ & $ n $ & $ n $ & $ \times $ & $ d $ & $ nz $ & $ n $ \\
\hline
Pass rate & 70/70 & 37/42 & 41/70 & 28/42 & 20/42 & 62/70 & 49/70 & 47/70
%\end{array}
%\end{equation*}
\label{tab:apple}
\end{longtable}
\end{footnotesize}

\subsection{ \texttt{REAL128} }

Flang 4.0 and PGI 16.3-0 do not support the \texttt{REAL128} real kind,
or any other extended precision kind.

In addition to the limitations detailed
in Sec. \ref{sec:res},
Oracle 12.6 does not support inverse hyperbolic intrinsics
for complex arguments of \texttt{REAL128} kind.

Although NAG 6.2 compiler does support \texttt{REAL128} kind,
it does not support IEEE arithmetic with it,
i.e. \texttt{IEEE\_SUPPORT\_DATATYPE(1.0\_REAL128)} is
false.

IBM XL Fortran 15.1.5 extended precision calculations
are enabled with
a non-standard \texttt{REAL(16)} kind, which does not
conform to IEEE binary128 format.
Therefore IBM Fortran does not support \texttt{REAL128} kind.

For \texttt{REAL128} real and complex variables
$ h \approx 1.2 \times 10^{4932} $,
$ t \approx 3.3 \times 10^{-4932} $
and
$ \epsilon \approx 1.9 \times 10^{-34} $.
The test results are summarised in Tab. \ref{tab:pear}.

\begin{footnotesize}
\begin{longtable} { l c c c c }
Test & GCC & Cray & Oracle & Intel \\
\hline
\endhead
\hline
\\
\caption{\normalsize Test results for \texttt{REAL128} kind.
$ t = $\texttt{ TINY(0.0\_REAL128)}; $ h = $\texttt{ HUGE(0.0\_REAL128)};
$ e = $\texttt{ EPSILON(0.0\_REAL128)}.}
\endfoot
$ \log( -h + \ic 0 ) $ &
 \ok & \ok & \ok & \ok \\
$ \log( -1 + \ic 0 ) $ &
 \ok & \ok & \ok & \ok \\
$ \log( -t + \ic 0 ) $ &
 \ok & \ok & \ok & \ok \\
$ \log( -t - \ic 0 ) $ &
 \ok & \ok & \ok & \ok \\
$ \log( -1 - \ic 0 ) $ &
 \ok & \ok & \ok & \ok \\
$ \log( -h - \ic 0 ) $ &
 \ok & \ok & \ok & \ok \\
\hline
$ \sqrt{ -h + \ic 0 } $ &
 \ok & \ok   & \ok & \ok \\
$ \sqrt{ -1 + \ic 0 } $ &
 \ok & \ok   & \ok & \ok \\
$ \sqrt{ -t + \ic 0 } $ &
 \ok & \ok   & \ok & \ok \\
$ \sqrt{ 0  + \ic 0 } $ &
 \ok & \ok   & \ok & \ok \\
$ \sqrt{ 0  - \ic 0 } $ &
 \ok & \ok   & \ok & \ok \\
$ \sqrt{ -t - \ic 0 } $ &
 \ok & \ok   & \ok & \ok \\
$ \sqrt{ -1 - \ic 0 } $ &
 \ok & \ok   & \ok & \ok \\
$ \sqrt{ -h - \ic 0 } $ &
 \ok & \ok   & \ok & \ok \\
\hline
$ \arcsin( - h + \ic 0 ) $ &
 $ o $ & $ o $ & $ \times $ & \ok \\
$ \arcsin( - 1 + \ic 0 ) $ &
 \ok   & \ok   & $ \times $ & \ok \\
$ \arcsin( - 1 - \ic 0 ) $ &
 \ok   & \ok   & $ \times $ & \ok \\
$ \arcsin( - h - \ic 0 ) $ &
 $ o $ & $ o $ & $ \times $ & \ok \\
$ \arcsin(   h + \ic 0 ) $ &
 $ o $ & $ o $ & $ \times $ & \ok \\
$ \arcsin(   1 + \ic 0 ) $ &
 \ok   & \ok   & $ \times $ & \ok \\
$ \arcsin(   1 - \ic 0 ) $ &
 \ok   & \ok   & $ \times $ & \ok \\
$ \arcsin(   h - \ic 0 ) $ &
 $ o $ & $ o $ & $ \times $ & \ok \\
\hline
$ \arccos( - h + \ic 0 ) $ &
 $ o $ & $ o $ & $ \times $ & \ok \\
$ \arccos( - 1 + \ic 0 ) $ &
 \ok   & \ok   & $ \times $ & \ok \\
$ \arccos( - 1 - \ic 0 ) $ &
 \ok   & \ok   & $ \times $ & \ok \\
$ \arccos( - h - \ic 0 ) $ &
 $ o $ & $ o $ & $ \times $ & \ok \\
$ \arccos(   h + \ic 0 ) $ &
 $ o $ & $ o $ & $ \times $ & \ok \\
$ \arccos(   1 + \ic 0 ) $ &
 \ok   & \ok   & $ \times $ & \ok \\
$ \arccos(   1 - \ic 0 ) $ &
 \ok   & \ok   & $ \times $ & \ok \\
$ \arccos(   h - \ic 0 ) $ &
 $ o $ & $ o $ & $ \times $ & \ok \\
\hline
$ \arctan( + 0 + \ic h ) $ &
 $ n $ & $ n $ & $ \times $ & $ d $ \\
$ \arctan( + 0 + \ic ( 1 + e ) ) $ &
 \ok   & \ok & $ \times $   & \ok \\
$ \arctan( + 0 + \ic 1 ) $ &
 \ok   & \ok & $ \times $   & \ok \\
$ \arctan( - 0 + \ic 1 ) $ &
 \ok   & \ok & $ \times $   & \ok \\
$ \arctan( - 0 + \ic ( 1 + e ) ) $ &
 \ok   & \ok & $ \times $   & \ok \\
$ \arctan( - 0 + \ic h ) $ &
 $ n $ & $ n $ & $ \times $ & $ d $ \\
$ \arctan( - 0 - \ic h ) $ &
 $ n $ & $ n $ & $ \times $ & $ d $ \\
$ \arctan( - 0 - \ic ( 1 + e ) ) $ &
 \ok   & \ok & $ \times $   & \ok \\
$ \arctan( - 0 - \ic 1 ) $ &
 \ok   & \ok & $ \times $   & \ok \\
$ \arctan( + 0 - \ic 1 ) $ &
 \ok   & \ok & $ \times $   & \ok \\
$ \arctan( + 0 - \ic ( 1 + e ) ) $ &
 \ok   & \ok & $ \times $   & \ok \\
$ \arctan( + 0 - \ic h ) $ &
 $ n $ & $ n $ & $ \times $ & $ d $ \\
\hline
$ \arcsinh ( + 0 + \ic h ) $ &
 $ o $ & $ o $ & $ \times $ & \ok \\
$ \arcsinh ( + 0 + \ic 1 ) $ &
 \ok   & \ok & $ \times $   & \ok \\
$ \arcsinh ( - 0 + \ic 1 ) $ &
 \ok   & \ok & $ \times $   & \ok \\
$ \arcsinh ( - 0 + \ic h ) $ &
 $ o $ & $ o $ & $ \times $ & \ok \\
$ \arcsinh ( + 0 - \ic h ) $ &
 $ o $ & $ o $ & $ \times $ & \ok \\
$ \arcsinh ( + 0 - \ic 1 ) $ &
 \ok   & \ok & $ \times $   & \ok \\
$ \arcsinh ( - 0 - \ic 1 ) $ &
 \ok   & \ok & $ \times $   & \ok \\
$ \arcsinh ( - 0 - \ic h ) $ &
 $ o $ & $ o $ & $ \times $ & \ok \\
\hline
$ \arccosh ( + 0 + \ic h ) $ &
 \ok & \ok & $ \times $ & \ok \\
$ \arccosh ( + 0 + \ic 1 ) $ &
 \ok & \ok & $ \times $ & \ok \\
$ \arccosh ( - 0 + \ic 1 ) $ &
 \ok & \ok & $ \times $ & \ok \\
$ \arccosh ( - 0 + \ic h ) $ &
 \ok & \ok & $ \times $ & \ok \\
$ \arccosh ( + 0 - \ic h ) $ &
 \ok & \ok & $ \times $ & \ok \\
$ \arccosh ( + 0 - \ic 1 ) $ &
 \ok & \ok & $ \times $ & \ok \\
$ \arccosh ( - 0 - \ic 1 ) $ &
 \ok & \ok & $ \times $ & \ok \\
$ \arccosh ( - 0 - \ic h ) $ &
 \ok & \ok & $ \times $ & \ok \\
\hline
$ \arctanh ( h     + \ic 0 ) $ &
 $ n $ & $ n $ & $ \times $ & $ d $ \\
$ \arctanh ( 1 + e + \ic 0 ) $ &
 \ok   & \ok & $ \times $   & \ok \\
$ \arctanh ( 1     + \ic 0 ) $ &
 \ok   & \ok & $ \times $   & \ok \\
$ \arctanh ( 1     - \ic 0 ) $ &
 \ok   & \ok & $ \times $   & \ok \\
$ \arctanh ( 1 + e - \ic 0 ) $ &
 \ok   & \ok & $ \times $   & \ok \\
$ \arctanh ( h     - \ic 0 ) $ &
 $ n $ & $ n $ & $ \times $ & $ d $ \\
$ \arctanh ( -h    + \ic 0 ) $ &
 $ n $ & $ n $ & $ \times $ & $ d $ \\
$ \arctanh ( -1 -e + \ic 0 ) $ &
 \ok   & \ok & $ \times $   & \ok \\
$ \arctanh ( - 1   + \ic 0 ) $ &
 \ok   & \ok & $ \times $   & \ok \\
$ \arctanh ( - 1   - \ic 0 ) $ &
 \ok   & \ok & $ \times $   & \ok \\
$ \arctanh ( -1 -e - \ic 0 ) $ &
 \ok   & \ok & $ \times $   & \ok \\
$ \arctanh ( - h   - \ic 0 ) $ &
 $ n $ & $ n $ & $ x $      & $ d $ \\
\hline
Pass rate & 50/70 & 50/70 & 14/14 & 62/70
%\end{array}
%\end{equation*}
\label{tab:pear}
\end{longtable}
\end{footnotesize}

\section{Recommendations for a future Fortran standard}
\label{sec:rec:log}

Fortran 2008 and the draft 2015 standards \cite{f2008,f2015}
prohibit \texttt{LOG} from accepting a zero argument, likely because
the imaginary part of $ \log ( \pm 0 \pm \ic 0 ) $ is undefined.

It is proposed to allow $ \log ( \pm 0 \pm \ic 0 ) $
with the following definition of the imaginary part:
\begin{equation}
\log ( 0 + \ic 0 ) = - \infty + \ic q
\qquad ;  \qquad
\log ( 0 - \ic 0 ) = - \infty - \ic q
\label{eq:newlog}
\end{equation}
where $ q > 0 $ is a processor dependent value.
A choice of $ q = \pi $ would be consistent with
$ \log $ values on the rest of the branch cut.
However, the exact value of $ q $ is immaterial.

Allowing $ \log ( \pm 0 \pm \ic 0 ) $
would be useful to the programmer, because it will
make the fundamental identity
\begin{equation}
z^a = \exp ( a \log z )
\label{eq:fox}
\end{equation}
valid for all $ z $.
An immediately useful example is $ \sqrt{ 0 \pm \ic 0 } $.
The proposed definition of $ \log ( \pm 0 \pm \ic 0 ) $ will
recover Eqn. \eqref{eq:sqrt} from App. \ref{sec:sqrt:derivation}:
\begin{eqnarray}
\sqrt{ 0 \pm \ic 0 }
& = & \exp( \frac{1}{2} \log ( 0 \pm \ic 0 ) )
 = \exp( \frac{1}{2} ( - \infty \pm \ic q ) )
\nonumber \\
& = & \exp( - \infty ) ( \cos \frac{ q }{ 2 } \pm \ic \sin \frac{ q } {2} )
 = 0 \pm \ic 0
\label{eq:moron}
\end{eqnarray}
where either the $ + $ or the $ - $ value of $ \pm $
is taken consistently. 

\section{Discussion}

Most compiler documentation referred to during
this work indicates that evaluation of
the 8 complex intrinsics is done via external calls,
typically to \texttt{libm}.
Therefore, the diversity of results between compilers, in Tabs. 
\ref{tab:monkey}, \ref{tab:apple} and \ref{tab:pear},
is surprising.
Although Cray and Oracle compilers show very similar
test failures, other compilers show different patterns.
This indicates that not all vendors use the same
algorithms and/or maths libraries.

As mentioned in the introduction,
both \texttt{LOG} and \texttt{SQRT} Fortran intrinsics accepted
complex arguments at least as far back as FORTRAN66,
and perhaps even earlier.
Therefore it was surprising to find that although
all compilers passed the \texttt{LOG} tests, multiple
\texttt{SQRT} failures was discovered in different compilers,
including overflow, underflow and wrong sign.
Given that all CPUs used in this work
fully support IEEE arithmetic
(see Tab. \ref{tab:compilers}) and
had hardware instructions for single
and double precision $ \sqrt{} $,
we speculate that the problems are likely 
in compiler implementations of complex $ \sqrt{} $,
e.g. Eqn. \eqref{eq:plane}.

Another surprise was that Flang did
significantly better in the tests than PGI
despite the fact that Flang front-end
is based on the PGI.

Many failures of type "$ n $", were obtained.
These are failures where NaN values were produced.
None of the 8 intrinsics should produce NaN results
on branch cuts, or indeed anywhere on the complex plane. 
Hence, such failures are obviously completely unacceptable.
This is the most obvious failure type, both to
the programmer and to the compiler or library developers.
The vendors should be able to find
and fix all such failures easily.

Another frequently observed failure type was
"$ o $", overflow, i.e. when $ \pm \infty $
results were produced
instead of the correct finite values.
These are most likely caused by overflow in intermediate
computations in the maths library.
These failures are more dangerous to
the programmer, because they can be hidden
by consecutive calculations.

In our opinion the most dangerous type of failure
to the programmer is type "$ s $", where the
sign of the real or the imaginary part of the result,
or both, is wrong.
Such failures will likely cause unexpected results
further down in the calculations, which will be very
hard to debug.
Expressions carefully derived in the Appendix
are intended to be used as a reference and a debugging aid.

Other failure types were seen less often.

Failure type "$ z $", where a zero result was obtained
instead of the correct non-zero normal value
was seen only together with other failure types,
overflow and NaN.
We therefore recommend the vendors to focus
on resolving failure types "$ n $" and "$ o $" first.

Failures of types
"$ d $",
where a subnormal result was
obtained while the processor did not support subnormals,
and
"$ m $",
where the magnitude of the real or the
imaginary part was clearly wrong,
were peculiar to a single vendor each.

It is important to emphasise that
only failures of type "$ n $", where NaN results
were produced,
can be interpreted as compiler non-conformance
with the standard.
This is because Fortran 2008,
or any previous Fortran standard,
requires very little in terms of accuracy
of floating point calculations.
Descriptions of many intrinsics have only the phrase
`processor-dependent approximation', e.g.
the result of $ \arccosh $(X) is defined
as `a value equal to a
processor-dependent approximation to the inverse hyperbolic
cosine function of X', where
`processor' is defined as a
`combination of a computing system and mechanism
by which programs are transformed for use on that computing
system' \cite{f2008}, i.e. it includes the compiler, the
libraries, but also the runtime environment
and the hardware.
Therefore, we interpret the test results only as
`quality of implementation'.

\section{Conclusions}

70 tests for complex Fortran 2008 intrinsics
\texttt{SQRT},
\texttt{LOG},
\texttt{ACOS},
\texttt{ASIN},
\texttt{ATAN},
\texttt{ACOSH},
\texttt{ASINH}
and
\texttt{ATANH}
on branch cuts
were designed for this work.
Only GCC and Intel Fortran compilers passed all
tests with complex arguments of kind \texttt{REAL32}
and only GCC passed all tests with complex arguments
of kind \texttt{REAL64}.
No compiler passed all tests with complex
arguments of kind \texttt{REAL128},
although the Intel compiler got very close.
Based on this limited testing, the user
is advised to deploy
inverse trigonometric and hyperbolic intrinsics,
$ \sqrt{} $ and $ \log $ on branch cuts with caution,
using extensive testing of the algorithms on known cases.
Unfortunately the need to use special code for calculations
on branch cuts has not yet disappeared completely.
We expect the quality of implementation in
all compilers to improve in line
with customer demands.
Finally, we welcome any feedback on our tests,
such as bug reports or results from other
compilers or compiler versions.
These can be submitted via \url{cmplx.sf.net}.

\section{Acknowledgments}

We acknowledge the use of several computational
facilities for this work:
The ARCHER UK National Supercomputing
Service (\url{http://www.archer.ac.uk});
Advanced Computing Research Centre of
The University of Bristol \\
(\url{http://www.bris.ac.uk/acrc})
and The STFC Hartree Centre \\
(\url{http://hartree.stfc.ac.uk}).
The STFC Hartree Centre is a research collaboration
in association with IBM providing High Performance
Computing platforms funded by the UK's investment
in e-Infrastructure.

\appendix

\section{Analytic solutions for 8 elementary complex functions on branch cuts}

This section contains brief but complete derivations
for the values of the 8 elementary complex functions studied in this work
on branch cuts.
The detailed derivations are given at
\texttt{cmplx.sf.net}.
The reader is referred to
NIST Handbook of Mathematical Functions (DLMF) \cite{dlmf} for all definitions.

The polar form, for $ z \ne 0 $:
\begin{equation}
z = | z | \exp \Arg z
\label{eq:oslik}
\end{equation}
where Arg$ z $ is defined in Tab. \ref{tab:babochka}
\cite[Sec. 1.9(i), Eqns. 1.9.5, 1.9.6]{dlmf}
and $ \omega $ is defined as follows:

\begin{equation}
\omega = \arctan \left| \frac{ y } { x } \right| ;
\qquad
\omega \in [ 0 , \pi / 2 ]
\label{eq:floor}
\end{equation}

\begin{table}[ht!]
\centering
\begin{tabular}{ c c c r }
\hline
quadrant & $ x $ & $ y $ & Arg$ z $ \\
\hline
1st & $ \ge 0 ; + 0 $ & $ \ge 0 ; + 0 $ & $ \omega $ \\
2nd & $ \le 0 ; - 0 $ & $ \ge 0 ; + 0 $ & $ \pi - \omega $ \\
3rd & $ \le 0 ; - 0 $ & $ \le 0 ; - 0 $ & $ - \pi + \omega $ \\
4th & $ \ge 0 ; + 0 $ & $ \le 0 ; - 0 $ & $ - \omega $ \\
\hline
\end{tabular}
\caption{Definition of Arg$ z $.}
\label{tab:babochka}
\end{table}

\subsection{ $ \log $ }

From \cite[Sec. 4.2(i), Eqn. 4.2.3]{dlmf}:

\begin{equation}
\log z = \log | z | + \ic \Arg z
\label{eq:chicken}
\end{equation}

$ \log z $ has a single branch cut
along the negative real axis, $ x \le 0 $
\cite[Sec. 4.2(i), Fig. 4.2.1]{dlmf}.

$ z = - a - \ic 0 , a > 0 $ is in the 3rd quadrant,
with $ \omega = 0 \Rightarrow $ Arg$ z = - \pi $.
$ | z | = a \Rightarrow \log | z | = \log a $. 
If $ 1 < a < + \infty $ then $ \log | z | > 0 $.
If $ 0 < a < 1 $ then $ \log | z | < 0 $.
If $ a = 1 $ then $ \log | z | = 0 $.

$ z = - a + \ic 0 , a > 0 $ is in the 2nd quadrant,
with $ \omega = 0 \Rightarrow $ Arg$ z = + \pi $.
The rest of the analysis follows the previous case.
The results are given in Tab. \ref{tab:log}.
See Sec. \ref{sec:rec:log} for the discussion of $ \log ( 0 \pm \ic 0 ) $.

\begin{table}[h!]
\centering
\begin{tabular}{ c c }
\hline
$ z $ & $ \log z $ \\
\hline
$ - a - \ic 0 $ & $ \log a - \ic \pi $ \\
$ - a + \ic 0 $ & $ \log a + \ic \pi $ \\
\hline
\end{tabular}
\caption{ $ \log z $ on the branch cut, $ a > 0 $.}
\label{tab:log}
\end{table}

\subsection{ $ \sqrt{} $ }
\label{sec:sqrt:derivation}

From \eqref{eq:oslik}:

\begin{equation}
\sqrt{ z }
  =
  \sqrt{ | z | } \exp \frac{ \Arg z } { 2 }
\label{eq:plane}
\end{equation}

$ \sqrt{ z } $ has a single branch cut along the negative
real axis, $ x \le 0 $, or $ \Arg z = \pm \pi $. 
For $ \Arg z = + \pi $, $ \sqrt{ z } $ is on
the positive imaginary axis.
For $ \Arg z = - \pi $, $ \sqrt{ z } $ is on
the negative imaginary axis, including $ | z | = 0 $:

\begin{equation}
\sqrt{ 0 + \ic 0 } = 0 + \ic 0 
; \qquad
\sqrt{ 0 - \ic 0 } = 0 - \ic 0 
\label{eq:sqrt}
\end{equation}
the sign of the real part is not important.

\subsection{ $ \arcsin $ }
\label{sec:arcsin}

From \cite[Sec. 4.23(iv), Eqn. 4.23.19]{dlmf}:

\begin{equation}
\arcsin z = - \ic \log ( \sqrt { 1 - z^2 } + \ic z )
\label{eq:arcsin}
\end{equation}

$ \arcsin z $ has 2 branch cuts, see Fig. \ref{fig:arcsin}.
Four cases are examined, one for each side of each branch cut.
In all cases $ a \ge 1 $ is a real value.

$ z = - a - \ic 0
\Rightarrow 
\ic z = \ic ( - a - \ic 0 ) = 0 - \ic a
\Rightarrow 
 z^2 = ( - a - \ic 0 ) ( - a - \ic 0 ) = a^2 + \ic 0
\Rightarrow 
 1 - z^2 = 1 - a^2 - \ic 0 
 \Rightarrow 
 \sqrt { 1 - z^2 } = \sqrt { 1 - a^2  - \ic 0 } $.
The expression under $ \sqrt {} $ is a complex number with
a non-positive real part and a negative zero imaginary part.
Therefore $ \sqrt { 1 - a^2 - \ic 0 } $ lies on the negative
imaginary axis.
This can be expressed as follows:
$ \sqrt { 1 - z^2 } = 0 - \ic \sqrt { a^2 - 1 } 
 \Rightarrow 
 \sqrt { 1 - z^2 } + \ic z
 = 0 - \ic ( \sqrt { a^2 - 1 } + a ) $.
The imaginary part of the last expression
is $ \le 0 $, therefore
$ \Arg ( \sqrt{ 1 - z^2 } + \ic z ) = - \pi / 2 
 \Rightarrow 
 \log ( \sqrt { 1 - z^2 } + \ic z )
 = \log ( \sqrt { a^2 - 1 } + a ) - \ic \pi /  2
 \Rightarrow 
 - \ic \log ( \sqrt { 1 - z^2 } + \ic z )
 = - \ic \log ( \sqrt { a^2 - 1 } + a ) - \pi / 2 
 = - \pi / 2 - \ic b $,
where
\begin{equation}
b = \log ( \sqrt { a^2 - 1 } + a ) \ge 0 .
\label{eq:b:sin}
\end{equation}
Hence
$ \arcsin z = - \pi / 2 - \ic b $.
This value lies in the 3rd quadrant.

$ z = - a + \ic 0 $ is analysed similar to the previous case.
$ \sqrt { 1 - z^2 } = 0 + \ic \sqrt { a^2 - 1 } 
 \Rightarrow 
 \sqrt { 1 - z^2 } + \ic z 
 = 0 + \ic ( \sqrt { a^2 - 1 } - a ) $,
where $ \sqrt{ a^2 - 1 } - a \le 0 $ and hence
$ | \sqrt{ a^2 - 1 } - a | = a - \sqrt{ a^2 - 1 } $.
Therefore $ \Arg ( \sqrt{ 1 - z^2 } + \ic z ) = - \pi / 2 $.
Since $ 0 < a - \sqrt{ a^2 - 1 } \le 1 $ then
$ \log ( a - \sqrt { a^2 - 1 } ) \le 0 $
and
$ \log ( \sqrt { 1 - z^2 } + \ic z ) =
   \log ( a - \sqrt { a^2 - 1 } ) - \ic \pi / 2 $.
Finally
$ - \ic \log ( \sqrt { 1 - z^2 } + \ic z ) 
 = - \ic \log ( a - \sqrt { a^2 - 1 } ) - \pi / 2
 = - \pi / 2 + \ic b $,
where $ b $ is given in Eqn. \eqref{eq:b:sin}.
Hence
$ \arcsin z = - \pi / 2 + \ic b $.
This value lies in the 2nd quadrant.

The identity
$ \arcsin ( - z ) = - \arcsin z $ 
from \cite[Sec. 4.23(iii), Eqn. 4.23.10]{dlmf}
is used to obtain expressions for $ \arcsin z $ for
the other sides of the branch cuts.
The results are summarised in Tab. \ref{tab:asin}.

\begin{table}[h!]
\begin{equation*}
\begin{array}{ r r }
\hline
\multicolumn{1}{c}{ z } & \multicolumn{1}{c}{ \arcsin z } \\
\hline
 - a - \ic 0 & - \pi / 2 - \ic b \\
 - a + \ic 0 & - \pi / 2 + \ic b \\
   a - \ic 0 & \pi / 2 - \ic b \\
   a + \ic 0 & \pi / 2 + \ic b \\
\hline
\end{array}
\end{equation*}
\caption{$ \arcsin z $ on branch cuts.
$ a \ge 1 $ and
$ b $ is given in Eqn. \eqref{eq:b:sin}.}
\label{tab:asin}
\end{table}

\subsection{ $ \arccos $ }

From \cite[Sec. 4.23(iv), Eqns. 4.23.19 and 4.23.22]{dlmf}
$ \arccos z = \pi / 2 - \arcsin z $.
$ \arccos z $ has the same 2 branch cuts as $ \arcsin z $,
hence we use the same 4 expressions for $ z $ as in
Sec. \ref{sec:arcsin}.
$ a $ and $ b $ are also as defined in Sec. \ref{sec:arcsin}.

$ z = - a - \ic 0 
 \Rightarrow 
 \arcsin z = - \pi / 2 - \ic b 
 \Rightarrow 
 \arccos z = \pi + \ic b $.
This value is in the 1st quadrant.

$ z = - a + \ic 0 
 \Rightarrow 
 \arcsin z = - \pi / 2 + \ic b 
 \Rightarrow 
 \arccos z = \pi - \ic b $.
This value is in the 4th quadrant.

Using the identity
$ \arccos ( - z ) = \pi - \arccos z $ from
\cite[Sec. 4.23(iii), Eqn. 4.23.11]{dlmf}
the values on the other sides of the branch cuts are obtained.
The results are summarised in Tab. \ref{tab:acos}.

\begin{table}[h!]
\begin{equation*}
\begin{array} { r r }
\hline
\multicolumn{1}{c}{ z } & \multicolumn{1}{c}{ \arccos z } \\
\hline
- a - \ic 0 & \pi + \ic b \\
- a + \ic 0 & \pi - \ic b \\
  a - \ic 0 &   0 + \ic b \\
  a + \ic 0 &   0 - \ic b \\
\hline
\end{array}
\end{equation*}
\caption{ $ \arccos z $ on the branch cuts.
$ a \ge 1 $
and
$ b $ is given in Eqn. \eqref{eq:b:sin}.}
\label{tab:acos}
\end{table}

\subsection{ $ \arctan $ }

$ \arctan z $ has 2 branch cuts along the imaginary axis,
$ y \ge 1 $ and $ y \le - 1 $, See Fig. \ref{fig:arctan}
and \cite[Sec. 4.23(ii), Fig. 4.23.1(ii)]{dlmf}.
The DLMF expression for $ \arctan $ in \cite[Sec. 4.23(iv), Eqn. 4.23.26]{dlmf}
has branch cuts along the real axis.
Expression in \cite[Sec. 4.23(iv), Eqn. 4.23.27]{dlmf}
can be used only for $ \arctan $ on branch cuts.
Hence we prefer to use the identity from \cite{kahan87}:
\begin{equation}
\arctan z = \frac{ \arctanh ( \ic z ) } { \ic }
\label{eq:kahan:atan}
\end{equation}
which leads to: 
\begin{equation}
\arctan z = \frac{ \ic }{ 2 } \log \frac{ 1 - \ic z } { 1 + \ic z }
 = \frac{ \ic }{ 2 } ( \log ( 1 - \ic z ) - \log ( 1 + \ic z ))
\label{eq:lesoval}
\end{equation}
which has the branch cuts along the imaginary axis.
Both sides of both branch cuts are analysed.
In the following $ a \ge 1 $ is a real number.

The 1st quadrant: $ z = 0 + \ic a 
 \Rightarrow 
\ic z = -a + \ic 0
 \Rightarrow 
1 - \ic z = a + 1 - \ic 0 
 \Rightarrow 
\log ( 1 - \ic z ) = \log ( a + 1 - \ic 0 ) $.
Arg$ ( a + 1 - \ic 0 ) = - 0
 \Rightarrow 
 \log ( 1 - \ic z ) = \log ( a + 1 ) - \ic 0 $.
Similarly
$ 1 + \ic z = 1 - a + \ic 0 
 \Rightarrow 
 \log ( 1 + \ic z ) = \log ( 1 - a + \ic 0 ) $.
Arg$ ( 1 - a + \ic 0 ) = \pi
 \Rightarrow 
 \log ( 1 + \ic z ) = \log ( a - 1 ) + \ic \pi 
 \Rightarrow 
 \log ( 1 - \ic z ) - \log ( 1 + \ic z )
 = \log ( a + 1 ) - \ic 0 - \log ( a - 1 ) - \ic \pi
  = \log ( a + 1 ) / ( a - 1 ) - \ic \pi $.
Finally
\begin{equation}
\arctan z
 = \frac{ \ic } { 2 }
   \left( \log \frac{ a + 1 } { a - 1 } - \ic \pi \right)
 = \frac{ \pi } { 2 } + \ic c
\end{equation}
where
\begin{equation}
c = \frac{ 1 } { 2 } \log \frac{ a + 1 } { a - 1 } \ge 0 .
\label{eq:c:arctan}
\end{equation}
If $ a = 1 \Rightarrow \Im ( \arctan z ) = + \infty $.
If $ a \to + \infty \Rightarrow \Im ( \arctan z ) \to + 0 $.

The 2nd quadrant: $ z = - 0 + \ic a 
 \Rightarrow 
\ic z = - \ic 0 - a 
 \Rightarrow 
1 - \ic z = a + 1 + \ic 0 
 \Rightarrow 
\log ( 1 - \ic z ) = \log ( a + 1 + \ic 0 ) $.
Arg$ ( a + 1 + \ic 0 ) = 0 
 \Rightarrow 
\log ( 1 - \ic z ) = \log ( a + 1 ) + \ic 0 $.
Similarly
$ 1 + \ic z = 1 - a - \ic 0 
 \Rightarrow 
\log ( 1 + \ic z ) = \log ( 1 - a - \ic 0 ) $.
Arg$ ( 1 - a - \ic 0 ) = - \pi
 \Rightarrow 
 \log ( 1 + \ic z ) = \log ( a - 1 ) - \ic \pi 
 \Rightarrow 
 \log ( 1 - \ic z ) - \log ( 1 + \ic z )
 = \log ( a + 1 ) + \ic 0 - \log ( a - 1 ) + \ic \pi
  = \log ( a + 1 ) / ( a - 1 ) + \ic \pi $.
Finally
\begin{equation}
\arctan z
 = \frac{ \ic } { 2 }
   \left( \log \frac{ a + 1 } { a - 1 } + \ic \pi \right)
 = - \frac{ \pi } { 2 } + \ic c 
\end{equation}
where $ c $ is as given in Eqn. \eqref{eq:c:arctan}.
If $ a = 1 \Rightarrow \Im ( \arctan z ) = + \infty $.
If $ a \to + \infty \Rightarrow \Im ( \arctan z ) \to + 0 $.

The identity
$ \arctan ( - z ) = - \arctan z $ 
from \cite[Sec. 4.23(iii), Eqn. 4.23.12]{dlmf}
is used to obtain expressions for $ \arctan z $ for
the values on the other sides of the branch cuts.
The results are summarised in Tab. \ref{tab:arctan}.

\begin{table}[!h]
\begin{equation*}
\begin{array} { r r }
\hline
\multicolumn{1}{c}{ z } & \multicolumn{1}{c}{ \arctan z } \\
\hline
 + 0 + \ic a &   \pi / 2 + \ic c \\
 - 0 + \ic a & - \pi / 2 + \ic c \\
 - 0 - \ic a & - \pi / 2 - \ic c \\
 + 0 - \ic a &   \pi / 2 - \ic c \\
\hline
\end{array}
\end{equation*}
\caption{ $ \arctan z $ on the branch cuts.
$ a \ge 1 $
and
$ c $ is given in Eqn. \eqref{eq:c:arctan}.}
\label{tab:arctan}
\end{table}

\subsection{ $ \arcsinh $ }

The most convenient expression for $ \arcsinh z $ is:
\begin{equation}
\arcsinh z = \ic \arcsin ( - \ic z )
\end{equation}
which can be obtained from Table 1 in \cite{kahan87}
or by combining \cite[Sec 4.37(iv), Eqn. 4.37.16]{dlmf}
with Eqn. \eqref{eq:arcsin}.
Accordingly the branch cuts are moved from the real
axis for $ \arcsin z $ to the imaginary axis for $ \arcsinh z $.
In the following $ a $ and $ b $ are as in Sec. \ref{sec:arcsin}.

$ z = - 0 - \ic a 
 \Rightarrow 
 - \ic z = - a + \ic 0 $.
From Tab. \ref{tab:asin}
$ \arcsin ( -a + \ic 0 ) = - \pi / 2 + \ic b
 \Rightarrow 
 \arcsinh z = - b - \ic \pi / 2 $.

$ z = 0 - \ic a 
 \Rightarrow 
 - \ic z = - a - \ic 0 $.
From Tab. \ref{tab:asin}
$ \arcsin ( - a - \ic 0 ) = - \pi / 2 - \ic b
 \Rightarrow 
\arcsinh z = b - \ic \pi / 2 $.

For the values on the other 2 sides of the branch cuts
we use the identity $ \arcsinh ( - z ) = - \arcsinh z $
from \cite[Sec. 4.37(iii), Eqn. 4.37.10]{dlmf}.
The results are summarised in Tab. \ref{tab:arcsinh}.

\begin{table}[h]
\begin{equation*}
\begin{array} { r r }
\hline
\multicolumn{1}{c}{ z } & \multicolumn{1}{c}{ \arcsinh z } \\
\hline
 + 0 + \ic a &   b + \ic \pi / 2 \\
 - 0 + \ic a & - b + \ic \pi / 2 \\
 + 0 - \ic a &   b - \ic \pi / 2 \\ 
 - 0 - \ic a & - b - \ic \pi / 2 \\
\hline
\end{array}
\end{equation*}
\caption{$ \arcsinh z $ on branch cuts.
$ a \ge 1 $
and
$ b $ is given in Eqn. \eqref{eq:b:sin}.}
\label{tab:arcsinh}
\end{table}

\subsection{ $ \arccosh $ }

From Table 1 in \cite{kahan87}, which is reproduced
in \cite[Sec. 4.37(iv), Eqn. 4.37.21]{dlmf}:
\begin{equation}
\arccosh z = 2 \log
 (  \sqrt{ ( z + 1 ) / 2 } + \sqrt{ ( z - 1 ) / 2 } )
\end{equation}
$ \arccosh z $ has a single branch cut along the real axis
at $ x \le 1 $.

$ z = - a + \ic 0 , a \ge 1 $
$ \Rightarrow $
$ \sqrt{ ( z + 1 ) / 2 } = \sqrt{ ( - a + 1 + \ic 0 ) / 2 } $
and
$ \sqrt{ ( z - 1 ) / 2 } = \sqrt{ ( - a - 1 + \ic 0 ) / 2 } $.
The real parts of both expressions under
$ \sqrt{} $ are $ \le 0 $.
The imaginary parts of both expressions under
$ \sqrt{ } $ are $ + 0 $,
i.e. the Arg of both expressions under
$ \sqrt{ } $ are $ + \pi $.
Hence the principal values of both square roots are on the
positive imaginary axis:
$ \sqrt{ ( z + 1 ) / 2 } = 0 + \ic \sqrt{ ( a - 1 ) / 2 } $
and
$ \sqrt{ ( z - 1 ) / 2 } = 0 + \ic \sqrt{ ( a + 1 ) / 2 } $
$ \Rightarrow $
$ \sqrt{ ( z + 1 ) / 2 } + \sqrt{ ( z - 1 ) / 2 }
  = 0 + \ic ( \sqrt{ ( a - 1 ) / 2 } + \sqrt{ ( a + 1 ) / 2 } ) $.
The imaginary part of the last expression is $ \ge 1 $,
therefore it is in the 1st quadrant.
Hence
$ \log
 ( \sqrt{ ( z + 1 ) / 2 } + \sqrt{ ( z - 1 ) / 2 } )
   = \log ( \sqrt{ ( a - 1 ) / 2 } + \sqrt{ ( a + 1 ) / 2 } )
     + \ic \pi / 2 $.
Further,
$ 2 \log ( \sqrt{ ( a - 1 ) / 2 } + \sqrt{ ( a + 1 ) / 2 } )
 = b $
$ \Rightarrow $
$ \arccosh z = b + \ic \pi $,
where $ b $ is given in Eqn. \eqref{eq:b:sin}

$ z = a + \ic 0 , -1 \le a \le 1 $
$ \Rightarrow $
$ \sqrt{ ( z + 1 ) / 2 } = \sqrt{ ( a + 1 + \ic 0 ) / 2 }
= \sqrt{ ( a + 1 ) / 2 } + \ic 0 $.
However in
$ \sqrt{ ( z - 1 ) / 2 } = \sqrt{ ( a - 1 + \ic 0 ) / 2 } $
the real and the imaginary parts of the expression
under $ \sqrt{ } $ are $ \le 0 $ and $ + 0 $ respectively,
meaning that the Arg
of this expression is $ + \pi $.
Hence the principal value of
$ \sqrt{ \ldots } $
is on the positive imaginary axis:
$ \sqrt{ ( z - 1 ) / 2 } = 0 + \ic \sqrt{ ( - a + 1 ) / 2 } $
$ \Rightarrow $
$ \sqrt{ ( z + 1 ) / 2 } + \sqrt{ ( z - 1 ) / 2 }
  = \sqrt{ ( a + 1 ) / 2 }
   + \ic \sqrt{ ( - a + 1 ) / 2 } $.
The absolute value of this expression is 1
and
$ \Arg ( \sqrt{ ( z + 1 ) / 2 } + \sqrt{ ( z - 1 ) / 2 } )
  = \arctan \sqrt{ ( -a + 1 ) / ( a + 1 ) } $.
Thus
\begin{equation}
\log \left(
 \sqrt{ \frac{ z + 1 } { 2 } } + \sqrt{ \frac{ z - 1 } { 2 } }
  \right)
   =
    0 + \ic \arctan \sqrt{ \frac{ -a + 1 } { a + 1 } }
\end{equation}
or $ \arccosh z = 0 + \ic d $
where
\begin{equation}
d = 2 \arctan \sqrt{ \frac{ -a + 1 } { a + 1 } }
\qquad ; \qquad
0 \le d \le \pi
\label{eq:e:arccosh}
\end{equation}
As $ a \to -1 \Rightarrow \Im ( \arccosh z ) \to \pi $.
As $ a \to 1 \Rightarrow \Im ( \arccosh z ) \to 0 $.
If $ a = 0 \Rightarrow \Im ( \arccosh z ) =  \pi / 2 $.

The case of
 $ z = a - \ic 0 , -1 \le a \le 1 $
is analysed similar to the case of
 $ z = a + \ic 0 , -1 \le a \le 1 $.
It is easy to show that
$ \sqrt{ ( z + 1 ) / 2 } + \sqrt{ ( z - 1 ) / 2 }
  = \sqrt{ ( a + 1 ) / 2 } - \ic \sqrt{ ( - a + 1 ) / 2 } $
and
$ \arccosh z = 0 - \ic d $,
where $ d $ is given in Eqn. \eqref{eq:e:arccosh}.
As $ a \to -1 \Rightarrow \Im ( \arccosh z ) \to - \pi $.
As $ a \to 1 \Rightarrow \Im ( \arccosh z ) \to 0 $.
If $ a = 0 \Rightarrow \Im ( \arccosh z ) = - \pi / 2 $.

Finally, the case of
$ z = - a - \ic 0 , a \ge 1 $
is analysed similar to the case of
$ z = - a + \ic 0 , a \ge 1 $.
The same logical steps lead to 
$ \sqrt{ ( z + 1 ) / 2 } + \sqrt{ ( z - 1 ) / 2 }
  = 0 - \ic ( \sqrt{ ( a - 1 ) / 2 } + \sqrt{ ( a + 1 ) / 2 } ) $
and
$ \arccosh z = b - \ic \pi $.
The results are summarised in Tab. \ref{tab:arccosh}.

\begin{table}[h]
\begin{equation*}
\begin{array} { r r c }
\hline
\multicolumn{1}{c}{ z } & \multicolumn{1}{c}{ a } & \arccosh z \\
\hline
- a + \ic 0 &          a \ge 1 & b + \ic \pi \\
  a + \ic 0 &  - 1 \le a \le 1 & 0 + \ic d \\
  a - \ic 0 &  - 1 \le a \le 1 & 0 - \ic d \\
- a - \ic 0 &          a \ge 1 & b - \ic \pi \\
\hline
\end{array}
\end{equation*}
\caption{ $ \arccosh z $ on the branch cut.
$ b $ and $ d $ are given in Eqns. \eqref{eq:b:sin}
and \eqref{eq:e:arccosh}.}
\label{tab:arccosh}
\end{table}

\subsection{ $ \arctanh z $ }

From Eqn. \eqref{eq:kahan:atan}
$ \arctanh z = \ic \arctan ( - \ic z ) $.
$ \arctanh z $ has 2 branch cuts along the real axis:
$ x \ge 1 $ and $ x \le -1 $.

$ z = a + \ic 0 , a \ge 1 
 \Rightarrow 
- \ic z = 0 - \ic a $.
From Tab. \ref{tab:arctan}
$ \arctan ( + 0 - \ic a ) = \pi / 2 - \ic c 
 \Rightarrow 
\arctanh z = c + \ic \pi / 2 $,
where $ c $ is given by Eqn. \eqref{eq:c:arctan}.

$ z = - a + \ic 0 , a \ge 1
 \Rightarrow
- \ic z = 0 + \ic a $.
From Tab. \ref{tab:arctan}
$ \arctan ( + 0 + \ic a ) = \pi / 2 + \ic c
 \Rightarrow
\arctanh z = -c + \ic \pi / 2 $,
where $ c $ is given by Eqn. \eqref{eq:c:arctan}.

Using the identity
$ \arctanh ( - z ) = - \arctanh ( z ) $ from
\cite[Sec. 4.37(iii), Eqn. 4.37.12]{dlmf},
the values on the other two sides of the branch cuts
are obtained.
The results are summarised in Tab. \ref{tab:arctanh}.

\begin{table}[h]
\begin{equation*}
\begin{array} { r r }
\hline
\multicolumn{1}{c}{ z } & \multicolumn{1}{c}{ \arctanh z } \\
\hline
  a + \ic 0 &    c + \ic \pi / 2 \\
- a + \ic 0 &  - c + \ic \pi / 2 \\
- a - \ic 0 &  - c - \ic \pi / 2 \\
  a - \ic 0 &    c - \ic \pi / 2 \\
\hline
\end{array}
\end{equation*}
\caption{ $ \arctanh z $ on the branch cuts.
$ a \ge 1 $
and
$ c $ is given in Eqn. \eqref{eq:c:arctan}.}
\label{tab:arctanh}
\end{table}

\subsection{ $ \log ( 2 h ) $ }
\label{sec:log2h}

Tabs. \ref{tab:asin}, \ref{tab:acos},
\ref{tab:arcsinh} and \ref{tab:arccosh} show that
on the branch cuts
$ \Im \arcsin z = \pm b $,
$ \Im \arccos z = \pm b $,
$ \Re \arcsinh z = \pm b $
and, on the part of the cut with $ x \le -1 $,
$ \Re \arccosh z = b $,
where $ b $ is given in Eqn. \eqref{eq:b:sin}.
The fact that the same expression for $ b $ appears
as either a real or an imaginary part in
these 4 complex functions on the branch cuts
is used in the tests.

When calculation is done using IEEE floating point arithmetic,
and $ a = h $, then
$ b = \log ( \sqrt{ h^2 - 1 } + h ) = \log ( 2 h ) $,
because within precision, $ p $, of
\texttt{REAL32}, \texttt{REAL64} or \texttt{REAL128}
real kinds, $ h + 1 = h - 1 = h $.
A truncated value of $ \log ( 2 h ) $,
denoted \texttt{log2h}, is used in the tests
for validating the calculated real or imaginary
parts of $ \arcsin $, $ \arccos $,
$ \arcsinh $ and $ \arccosh $:

\begin{lstlisting}
log2h=real(int(log(2.0_fk)+log(huge(0.0_fk))),kind=fk)
\end{lstlisting}
where \texttt{fk} is either \texttt{REAL32},
\texttt{REAL64} or \texttt{REAL128}.

The calculated difference
$ \log ( 2 h ) - \texttt{log2h} $
is much greater than the expected relative error
in calculated $ \log ( 2 h ) $,
see Sec. \ref{sec:intro}.
Hence tests can be constructed to
require that the following is true:
$ \Im \arccosh ( - h + \ic 0 ) > \texttt{log2h} $,
with similar tests for the other 3 functions.
Failures of such tests are assigned type "$ m $",
see Tab. \ref{tab:entries}.

\bibliography{ref}
\bibliographystyle{unsrt}

\end{document}